\documentclass[letterpaper]{article} 
\usepackage[]{aaai2026}  
\usepackage{times}  
\usepackage{helvet}  
\usepackage{courier}  
\usepackage[hyphens]{url}  
\usepackage{graphicx} 
\usepackage{natbib}  
\usepackage{caption} 
\usepackage{algorithm}
\usepackage{algorithmic}

\usepackage{newfloat}
\usepackage{listings}
\DeclareCaptionStyle{ruled}{labelfont=normalfont,labelsep=colon,strut=off} 
\floatstyle{ruled}
\newfloat{listing}{tb}{lst}{}
\floatname{listing}{Listing}
\usepackage{supertabular}
\usepackage{color}
\usepackage{tabularray}
\usepackage{dsfont}
\usepackage{amsmath}
\usepackage{booktabs}
\usepackage{multirow}
\usepackage{longtable}
\usepackage{siunitx}
\usepackage{array}
\definecolor{gray}{HTML}{808080}
\definecolor{teal}{HTML}{21908C}
\definecolor{yellow}{HTML}{FDE725}
\definecolor{blue}{HTML}{1E2BC5}
\definecolor{purple}{HTML}{5D3FD3}
\definecolor{green}{HTML}{238E12}

\title{Context and Symmetry in Auditing:\\A Case Study of Skeleton Inference in Motion Capture}
\author {
    Emma Harvey\equalcontrib\textsuperscript{\rm 1},
    Emanuel Moss\equalcontrib\textsuperscript{\rm 2, 3},
    Hauke Sandhaus\textsuperscript{\rm 1},
    Abigail Z.\ Jacobs\textsuperscript{\rm 4},
    Mona Sloane\textsuperscript{\rm 2}
}
\affiliations {
    \textsuperscript{\rm 1}Cornell Tech, \textsuperscript{\rm 2}University of Virginia, 
    \textsuperscript{\rm 3}Intel Corporation,
    \textsuperscript{\rm 4}University of Michigan\\
    evh29@cornell.edu, emanuel.moss@intel.com, hgs52@cornell.edu, azjacobs@umich.edu, mona.sloane@virginia.edu
}

\begin{document}

\maketitle

\defcitealias{HHS}{DHHS 1994}

\begin{abstract}
Humans are increasingly expected to interact with AI systems that observe and make inferences about them---but do these systems actually work? A standard approach to answering this question is AI auditing. Conducting an AI audit requires identifying how a system behaves (i.e., determining what types of inputs to audit it with and then observing and documenting actual system behavior) and contrasting that with how a system should behave (i.e., determining what the nominal outputs of a system should look like). We argue that this is best done through a \textbf{contextual audit}, which we introduce as a method for auditing measurements within the context of the practices that produce them. We show how contextual auditing enables the interrogation of assumptions implicit in the audit process and allows auditors to be explicit about what serves as \textit{ground truth}, which we define as verifiable measurements about the real world against which systems are evaluated. We outline how the concept of \textbf{symmetry} from science and technology studies can enable audits when ground truth is unknown, unknowable, or contested. Finally, to demonstrate how contextual and symmetric audits can be conducted in practice, we present a case study of skeleton inference in motion capture and propose areas suggested by our case study as particularly fruitful for future audits of motion capture systems.\looseness=-1
\end{abstract}

\section{Introduction}\label{sec:1-introduction}
Humans are increasingly enrolled into interactions with AI systems that observe and make inferences about them. These systems, which can be thought of as making \textit{measurements} about people's bodies or behaviors, now moderate high-stakes outcomes like employment (e.g., candidate assessments~\cite{landers2022game}) and physical health (e.g., workplace safety monitoring~\cite{pr_xsens}). Despite the widespread adoption of AI systems, a crucial question is often left unanswered: do they work as intended~\cite{raji_fallacy_2022}? A standard approach to answering this question relies on \textit{AI auditing}, which typically entails comparing the \textit{actual} outputs of a system to \textit{nominal} expected outputs---in other words, comparing how the system does behave to how it should behave~\cite{birhane_ai_2024, mei_addressing_2026, sandvig_auditing_2014, sloane_assessing_2023}. This immediately raises two questions. First, \textbf{what inputs are relevant for identifying patterns in a system's actual outputs?} Second, \textbf{what should a system's nominal outputs look like}~\cite{mei_addressing_2026}\textbf{?}\looseness=-1

A typical approach to AI auditing is to elicit outputs from a system in a systematic way to capture a specific behavior \cite{bandy_problematic_2021}. Researchers often use pre-existing or synthetic data as inputs to produce outputs that can be analyzed to assess how well they match nominal outputs~\cite[e.g.,][]{fleisig_linguistic_2024, hofmann_ai_2024, imana_external_2025, wilson_gender_2024}. However, such approaches have been criticized for failing to capture real-world use, and thus the potential real-world impacts, of systems~\cite{harvey_framework_2025, wang_inadequacy_2025}. Instead, drawing on \citet{sloane2023introducing}, we argue that what is required is an auditing approach that uses inputs and nominal outputs that are meaningful within the contexts in which AI systems are actually used. To address this, we introduce the \textbf{contextual audit}, which makes claims about the accuracy of system outputs \textit{within the context of the practices that produce them}. In this work, we lay out the importance of contextual audits and their role within the broader domain of AI auditing. Then, to demonstrate how contextual audits can be conducted, we present a case study of skeleton inference in motion capture (mocap). \textit{Skeleton inference}, a basic task in mocap, involves sensing landmarks on the surface of a body and applying algorithms to infer the position and motion of skeletal elements within the body. Skeleton inference can be used to produce \textit{body segment parameters} (BSPs), which are measurements of various properties (e.g., length) of body segments (e.g., forearm). BSPs are then used in a variety of mocap applications, such as workplace safety monitoring and gait recognition.\looseness=-1

Like other forms of audits, contextual audits make claims about the behavior of a system by comparing its actual outputs to \textit{nominal outputs}, which describe ideal behavior. A typical approach requires that a system meet some threshold for accuracy (overall or in terms of relative accuracy, e.g., across demographic groups). Determining accuracy requires comparing system outputs to a \textit{ground truth}, which we define as a verifiable set of measurements about the real world. To conduct a contextual audit of skeleton inference, we sought to assess the accuracy of BSPs produced by a mocap system. However, there is no single agreed-upon system of \textit{ground truthing} (i.e., determining what should be taken as ground truth) in the context of BSPs. For example, should the length of a forearm be measured based on the length of the radius or ulna (neither of which are directly observable without dismemberment), or should it be based on the distance between landmarks on the surface of the body (which may change depending on forearm position or the presence of soft tissue)? Deciding what accuracy means is far from simple---but it is crucial for AI auditing. In our case study audit, we compare BSPs derived from mocap systems to those derived using tape measured anthropometric techniques. However, anthropometry, like mocap, is a practice of measurement that attempts to establish, but cannot directly measure, ground truth.\looseness=-1 

To address this, we propose that contextual audits dealing with the measurement of concepts where ground truth is unknown, unknowable, or contested must borrow from the science and technology studies (STS) concept of \textbf{symmetry}~\cite{bloor1991knowledge}. Symmetry requires treating competing theories of truth with the same analytical approach---as if either might be true or false---and draws attention to the social practices of establishing their validity \cite{jasanoff2019controversy}. We argue that symmetry can be adapted for AI audits: it guides auditors to treat the outputs of each measurement modality as provisionally true, and then to interrogate the implications of doing so. Because assessing implications of systems' behaviors is the focus of AI auditing, symmetry enables a practice for assessing AI systems even in the absence of a ground truth. Practically, symmetry can be used to explore where measurements produced through different modalities are in tension and not easily reconcilable over time and then to point to where the assumptions underpinning a system should be further investigated. Overall, we make the following contributions:\looseness=-1
\begin{enumerate}
    \item We introduce \textbf{contextual audits} for auditing measurements in the context of the practices that produce them, and we show how they enable interrogating what serves as ground truth and identifying the consequences of using a given ground truth to audit an AI system.
    \item We show how \textbf{symmetry} can enable audits when ground truth is unknown, unknowable, or contested.
    \item We demonstrate the applicability of contextual auditing through a case study audit of skeleton inference in motion capture.
\end{enumerate}
\section{Background and Related Work}\label{sec:2-background}
In this section, we briefly describe existing approaches to AI auditing (\S\ref{subsec:background:ai-auditing}) and define the concepts from sociology and STS---social practice theory and symmetry---that underlie our work (\S\ref{subsec:background:sociology-STS}). To ground our case study, we also discuss the practice of measurement in general and of skeleton inference in mocap in particular (\S\ref{subsec:background:measurement}).

\subsection{AI Auditing}\label{subsec:background:ai-auditing}
AI auditing is the practice of comparing how AI systems do behave to how they should behave, with the goals of identifying harms, creating accountability, or improving systems~\cite{bandy_problematic_2021, birhane_ai_2024, becerrasandoval_historical_2025, mei_addressing_2026, mokander2023auditing, raji_closing_2020, rhea2022external, sloane_algorithmic_2021, sloane_silicon_2022, sloane_assessing_2023}. \citet{sandvig_auditing_2014} emphasize that system behavior should be elicited through ``a combination of [a system] and its data,'' i.e., by querying a system with specific inputs and observing its outputs~\cite{rhea2022external}. This formulation of AI auditing raises two key questions. First, how do we determine how a system actually behaves---what inputs are relevant
for identifying patterns in a system's actual outputs~\cite{mei_addressing_2026}? Second, how do we
decide how a system should behave---what should a system’s nominal outputs look like? Answering these questions requires considering:\looseness=-1
\begin{enumerate}
    \item The \textit{context} in which AI systems are designed and used.
    \item The \textit{ground truth} according to which AI systems can be evaluated---ground truth is taken as nominal outputs.
\end{enumerate}

\subsubsection{Context in AI Auditing.} Deciding what inputs to use to conduct an AI audit is a non-trivial question. Collecting real-world input data is time-consuming and expensive, and can raise user privacy concerns (e.g., if users share personal information as a part of real-world system use). As a result, auditors often adapt pre-existing data or create their own synthetic data~\cite[e.g.,][]{fleisig_linguistic_2024, hofmann_ai_2024, wilson_gender_2024}. Then, audits are conducted by examining the outputs produced by using those inputs to prompt the AI system. However, this approach is inherently limited: it does not allow for the evaluation of AI systems in their normal context of use. \citet{harvey_framework_2025} argue that audits should consider \textit{ecological validity}, or ``the extent to which experimental findings can generalize to the `real world' situation that a researcher wishes to understand'' \cite{Kihlstrom_ecological_2021}. This means that the inputs should be produced as they would be produced in a real-world application, and outputs should be analyzed according to their actual context of use. \citet{wang_inadequacy_2025} underscore the importance of contextualized inputs: they find that the performance of large language models (LLMs) on widely-used benchmarks depends on the chat history of the prompter, meaning that generic evaluations do not necessarily provide an understanding of performance and impact for real-world users. Looking at outputs of a system divorced from their context of use also provides a misleading understanding of system performance~\cite{gosciak2026llmssocialservicesdoes}. For example, \citet{pruss_ghosting_2023} finds that judges rarely use a recidivism prediction tool, rendering audits of it meaningless; similarly, \citet{cheng_how_2022} find that audits of a child welfare assessment tool overstate racial bias because caseworkers' selective adoption of the tool mitigates some biases. In this work, we propose the \textbf{contextual audit} as an approach for conducting AI audits that consider how systems are actually used in practice. In a contextual audit, inputs are produced following actual practices of system users, and outputs are evaluated in the context of their use.\looseness=-1

\subsubsection{Ground Truth in AI Auditing.}
An AI audit requires a comparison between nominal and actual behaviors~\cite{sloane_assessing_2023}. Researchers often test normative expectations, e.g., that systems do not encode ``harmful discrimination''~\cite{sandvig_auditing_2014}. A common operationalization of harmful discrimination is whether systems are equally accurate (i.e., produce actual outputs that are equally close to nominal ones) across demographic groups~\cite{compas, buolamwini_gender_2018, obermeyer_dissecting_2019, wilson_building_2021}. For many systems, however, determining accuracy is challenging, or even impossible: ground truth is not directly observable either because it is a theoretical (and potentially contested) concept (e.g., creditworthiness) or because it simply cannot be directly accessed (e.g., certain anatomical measurements)~\cite{jacobs_measurement_2021}. In other words, there is no agreed upon ground truth data---or nominal value---according to which a system can be evaluated. In these cases, it is not immediately clear how to conduct an AI audit. In the absence of ground truth, researchers have tested whether AI systems achieve parity (i.e., produce similar outputs across demographic groups)~\cite{sweeney_discrimination_2013, feng_has_2022, wilson_building_2021} or reliability (i.e., produce repeatable outputs)~\cite{rhea_resume_2022, rhea2022external}. However, neither approach fully answers the question of whether AI systems---specifically, systems that make measurements about people---produce outputs that are \textit{valid}, i.e., that meaningfully measure what they are intended to measure~\cite{jacobs_measurement_2021}. In this work, we propose that the concept of \textbf{symmetry}, drawn from STS, can be a powerful tool for enabling audits in the absence of an agreed upon ground truth or in situations in which multiple ground truths are conflicting or contested~\cite{bloor1991knowledge}.\looseness=-1

\subsection{Sociology and STS Concepts for AI Auditing}\label{subsec:background:sociology-STS}
Since its inception, AI auditing has drawn from multiple disciplines. It was originally proposed as an adaptation of social science auditing, which uses empirical methods like field experiments to identify and quantify discrimination in societal systems~\cite{bertrand_emily_2004, sandvig_auditing_2014, vecchione_algorithmic_2021}. As AI audits become increasingly important for ensuring AI systems can safely be deployed in society, researchers and practitioners have drawn lessons from other fields, including financial auditing and engineering or medical safety evaluations~\cite{mokander2023auditing, raji_closing_2020, rismani_plane_2023}. We demonstrate that AI auditing can further benefit from drawing more explicitly on additional approaches from the fields of sociology and STS.\looseness=-1

\subsubsection{Social Practice Theory.} Contextual audits require an understanding of the practices by which inputs to an AI system are produced and outputs of an AI system are used. \textit{Social practice theory} is an approach to examining how technical systems operate within social settings that focuses on the activities and behaviors that individuals and groups conduct when engaging with a technological object. Rather than focus on the object or the individual, social practice theory focuses on the practices that stabilize, change, and re-stabilize over time. Social practices are composed of three elements: \textit{materials} (objects), \textit{competences} (skills and knowledge), and \textit{meanings} (understandings and beliefs) that change and reassemble in specific contexts~\cite{shove_design_2007, shove_dynamics_2012}. Social practice theory can provide a frame for understanding the contexts in which AI systems are designed and used, particularly for understanding how AI systems operate as novel materials being incorporated into already-existing sets of competences and meanings~\cite{harvey-cadaver, sloane_introducing_2022}. \looseness=-1

\subsubsection{Symmetry.} 
When concepts can be measured in multiple ways~\cite{hand2004measurement, klein1988science}, \textit{concurrent validity} assessments are often conducted, in which one measurement model (the particular systematization and operationalization of the underlying concept) is evaluated against another~\cite{jacobs_measurement_2021}. This raises a question: which model is the better approximation of the underlying concept? While one model can be treated as the basis for comparison with another model, either (or both) may be flawed. Treating one or the other as ground truth obscures the shortcomings of that model while also holding the other to an impossible---and potentially flawed---standard. Here, we draw upon the general principle of \textit{symmetry}, which was originally developed for the social study of knowledge and widely deployed in STS~\cite{bloor1991knowledge, jasanoff2019controversy}. For STS scholars, the impulse to separate `true' scientific descriptions of reality from `false' ones is held at bay by applying a symmetrical lens to each. Each scientific description is analyzed using the same techniques, as if it has the same likelihood of being true or false. In STS, symmetry is deployed not to erase the distinction between truth and error~\cite{fuller2016embrace}, but to better understand how scientific claims are made and come to be accepted by a broader community. Similarly, we argue that AI auditing can adopt symmetry to examine different measurement modalities without needing to hold one as a ground truth against which the other is compared. Rather, symmetry can help analyze the social and technical practices through which systems come to be accepted as valid and reliable.\looseness=-1

Symmetry is important for AI audits because it acknowledges the often-contingent nature of ground truth and exposes the assumptions needed to accept data as ground truth. This is crucial for identifying potential gaps between underlying concepts and how they are represented as ground truth data. If such gaps go undetected by an audit process, audits can fail to detect system harms. For example, toxicity benchmarks may produce dramatically different results depending on the populations and guidelines used to produce the benchmarks~\cite{fleisig_when_2023, kirk2024the}. Holding a symmetrical view in which each benchmark \textit{could} represent ground truth for a given context allows auditors to unpack the assumptions underlying the data in constructive ways; e.g., to justify using one benchmark instead of another, to modify or improve a benchmarking practice, or to buttress an audit with additional investigative techniques. Crucially, symmetry frames its analysis within the context to which the audit tool is being applied by asking how the assumptions that have been made in the design of a system affect a specific implementation in a particular place, at a particular time.\looseness=-1

\subsection{Measurement and Skeleton Inference}\label{subsec:background:measurement}
Measurement is the quantification of an (often unobservable) underlying concept using observable data~\cite{bandalos2018measurement}. Measurement is distinct from prediction and related tasks like generation in that measurement is an attempt to approximate an existing ground truth~\cite{mussgnug_predictive_2022}.\looseness=-1

Mocap systems are AI systems that perform measurement tasks. They consist of sensors, which collect data about surfaces of bodies, and algorithms, which make inferences based on that data. Among the most basic tasks conducted by mocap systems is \textit{skeleton inference}, which seeks to measure the position and motion of skeletal elements within a body.\footnote{We refer to \textit{skeletal elements} rather than \textit{bones} to indicate that skeleton inference infers the position and motion of rigid structures within the body, but that these structures are typically simplified models of skeletons.\looseness=-1} Skeleton inference produces BSPs, which measure various properties of body segments. Skeleton inference and BSPs are the building blocks for a variety of mocap applications, ranging from low-stakes to high-stakes settings. For example, they enable the creation of lifelike human motion for video games. They are also used in biometric tasks like gait recognition, which seeks to identify individuals based on how they walk, and for health-related tasks, including workplace safety monitoring and physical therapy~\cite{HAN2013131, topham_human_2022}. As mocap applications proliferate, it becomes increasingly important to understand how these technologies perform across the wide range of human bodies that they might come into contact with.\looseness=-1

There is good reason to believe that skeleton inference may encode discriminatory biases by performing differently well (and, in turn, producing differently accurate BSPs) for different types of bodies. Modern mocap systems and the inferences they make are largely built on data from a small number of thin male cadavers~\cite{harvey-cadaver}. Prior work has found that inferences made by mocap systems do not generalize equally well to individuals whose bodies do not match those in the data~\cite{durkin_analysis_2003}. Importantly, skeleton inference is challenging to ``ground truth'': the nominal values of BSPs produced through skeleton inference are often unknown or contested~\cite{bodenheimer1997process}. Concurrent validity assessments are often used to evaluate mocap systems that do skeleton inference---but \citet{harvey-cadaver} have found that such assessments often rest on strong assumptions and may erroneously treat flawed measurements as ground truth.\looseness=-1

In this work, we conduct a case study audit of the BSPs produced by a mocap system. Beginning with a straightforward concurrent validity assessment, in which we compare BSPs produced by a mocap system to those produced through an alternative modality (anthropometric measurements taken with a tape measure), we show how context and symmetry are critical tools for auditors seeking to evaluate real-world performance of AI systems without access to ground truth data.\looseness=-1
\section{Framework for Contextual Audits}\label{sec:3-framework}
A \textbf{contextual audit} makes claims about outputs of AI systems \textit{within the context of the practices that produce them}. This means that contextual audits must begin by selecting a specific application of a system---they cannot be expected to evaluate every set of outputs that a system is capable of producing. Then, contextual audits involve identifying the particular social practices that are used to produce outputs in the selected application. In this way, contextual audits help to answer the question, ``what inputs are relevant for identifying patterns in a system's actual outputs?'' by suggesting that we follow the practices specific to our desired context of use. Contextual audits also surface relevant sets of assumptions underpinning the application of the AI system. These assumptions can point to \textit{facets}, or features across which system outputs should be compared, as well as ideals describing how systems should behave across those facets. In this way, contextual audits help to answer the question, ``what should a system's nominal outputs look like?''~\cite{rhea_resume_2022}.\looseness=-1

While the standard approach to AI auditing centers quantitative analysis, contextual audits fundamentally require qualitative research. Understanding the practices that produce the outputs of AI systems requires deep engagement with the users who conduct those practices. A contextual audit should therefore include a field observation or an ethnography of the users who operate an AI system within a specific context; it may also include informational interviews with those users. The context in which users operate is also shaped by the AI system itself. In particular, historical decisions about how systems should be built and documentation describing how systems should be used are both factors that may be taken for granted by users, but that deeply affect their context of use. Thus, contextual audits should also include a review of the research, data, and design decisions underpinning an AI system, as well as the system's documentation. Appropriate methods for this aspect of a contextual audit may include a systematic literature review~\cite[e.g.,][]{harvey-cadaver}, a historical analysis~\cite{becerrasandoval_historical_2025}, or a trace ethnography~\cite{geiger_trace_2011}.\looseness=-1

To support this work, we adapt the socio-technical matrix proposed by \citet{sloane_silicon_2022} and extended by \citet{rhea_resume_2022} and \citet{sloane2025making}. That matrix, originally designed to facilitate audits of automated decision-making systems used in hiring, prompts auditors to examine the context in which a system is intended to operate, interrogate how the system works in practice, and identify the assumptions and epistemological roots that underpin the system (and thus what is considered its ground truth). The matrix, which we have further adapted here, aids in identifying assumptions about what can and should be accepted as nominal values by helping auditors define, from a system's many possible purposes, the specific goals that it is being used to achieve. Auditors can then isolate the sets of assumptions that underpin those goals and trace how those assumptions are operationalized in specific measurements produced by the system. The matrix, then, enables auditors to audit a complex AI system by clearly stipulating the specific purpose it is being used for, the capability drawn on to accomplish that purpose, the set of assumptions that underpin that capability, and the measurements that are both expected by that capability and actually produced by it. We adapt the socio-technical matrix to focus more generally on measurement systems and propose that contextual audits should use the qualitative methods described above to answer the following before conducting quantitative testing:\looseness=-1

\begin{enumerate}
    \item Identify the object(s) of capture and inference
    \item Determine how objects are made legible to the system 
    \item Examine how ground truth is and has historically been established
    \item Determine what assumptions ``ground truths'' encode
    \item Consider how these assumptions may produce harms when operationalized in the relevant context of use
\end{enumerate}
\section{Case Study Audit: Skeleton Inference}\label{sec:4-case-study}
Mocap systems operate in multiple contexts, ranging from entertainment to health and safety to biometric surveillance~\cite{mundermann2006evolution, berner2020concurrent}. Each context brings with it a unique and specific set of practices through which measurements are produced, and each practice is defined by its own set of materials, competences and meanings. Our case study audit focuses on the context of \textbf{mocap for creative design}, which entails capturing bodies in motion as inputs for creative projects~\cite{dykes2009towards}. This can include faithfully recording body movement (e.g., for capturing an athlete's signature style for a video game) as well as distorting bodies (e.g., for enabling CGI of nonhuman bodies performing natural-looking movements). When bodies are recorded, mocap systems perform skeleton inference and produce BSPs, which may then be used in downstream tasks.\looseness=-1

\paragraph{Audit Stakes.} Mocap for creative design is relatively low-stakes when compared to other mocap use cases, such as workplace safety monitoring, biometric surveillance, and medical diagnosis: a person's physical health and safety are not at risk if they are mismeasured by a mocap system in a creative design task. Nevertheless, even a creative design mocap system has real stakes. In particular, mocap systems can contribute to representational harms if they misrepresent or erase a particular body type in entertainment settings~\cite{blodgett_language_2020}. Further, as \citet{harvey-cadaver} show, materials, competences, or meanings initially developed for entertainment are often adapted for higher-stakes settings, meaning harms can travel with them. For example, the Microsoft Kinect mocap system was originally marketed as a video game controller, but quickly became the basis for home health and rehabilitation mocap applications due to its low price and small size~\cite{blumrosen2016real, su2014kinect}. Thus, while our audit focuses on mocap for creative design, its results point to areas that should be scrutinized in future audits of higher-stakes mocap systems.\looseness=-1

\paragraph{Audit Setting.} In our audit, we worked with mocap practitioners at New York University (NYU) whose work focused on researching augmented, virtual, and extended reality as well as working with external clients, including companies, museums, and artists, to produce creative outputs.\footnote{\url{https://engineering.nyu.edu/research/nyu-tandon-yard}} Mocap at NYU is conducted on a \textit{mocap stage}, a 40'x35' platform that is surrounded by an OptiTrack Prime 13 mocap system, which includes 24 cameras arranged around the stage.\footnote{\url{https://optitrack.com/}} The OptiTrack Prime 13 is a marker-based system: it uses infrared cameras to track markers placed on the surface of the human body. Those markers are then used to infer the position and motion of the underlying skeletal elements based on one of several \textit{body models}, which are mathematical models built on historical data that come with OptiTrack's Motive software. The body models perform skeleton inference and produce BSPs. Our study was approved by the NYU Institutional Review Board (IRB-FY2023-7677).\looseness=-1

\subsection{Incorporating Context}\label{subsec:case-study:context}

\paragraph{Qualitative Methods Reveal Context.} To understand the context of mocap for creative design, we conducted field work directly with mocap practitioners at NYU in June and July 2023. We first spent a day in training. This included shadowing the practitioners as they conducted mocap and learning how to conduct mocap activities ourselves. We were trained on how to calibrate cameras,\footnote{Calibration consists of waving a bar with markers attached at fixed points across the mocap stage to allowed cameras to triangulate their positions relative to each other.\looseness=-1} how to identify body landmarks and affix markers to them, how to capture motion using the OptiTrack system, and how to troubleshoot the system. Importantly, practitioners walked us through the actual practices that they followed as opposed to potential best practices that might exist in academic literature or system documentation. We also engaged practitioners in question-and-answer sessions to document their understandings of the mocap technology and the relevance of each of their techniques to the overall practice they were engaged in. Following the day of training, we conducted two (non-consecutive) weeks of field experiments using the mocap system, for which we recruited 24 participants to have their BSPs measured. During these experiments, we asked the mocap practitioners for guidance or help debugging the mocap system as needed. Finally, we conducted informal interviews with the mocap practitioners to better understand how they perceived the materials, competences, and meanings that were central to their practice of mocap. Throughout, we took extensive field notes to capture both the practices that practitioners described as well as how we applied and adapted those practices ourselves.\looseness=-1

We supplemented our field work with a close examination of the OptiTrack documentation and a historical literature review~\cite[see][]{harvey-cadaver}. Together, the documentation and literature review allowed us to uncover materials, competences, and meanings that have been baked into the OptiTrack system over time, and thus help constitute the context in which the mocap practitioners we observed operate.\looseness=-1

\begin{table*}[ht!]
\small
\centering
\begin{tabular}{>{\hspace{0pt}}m{0.2\linewidth}>{\hspace{0pt}}m{0.38\linewidth}>{\hspace{0pt}}m{0.38\linewidth}} 
\toprule
 & \textbf{Marker-Based Motion Capture} & \textbf{Tape Measure-Based Anthropometry} \\
 \midrule
\textbf{Identify the object(s) of capture and inference} & 
\textit{Capture}: Markers on bodies \par{}\vspace{3pt}
\textit{Inference}: Position and motion of skeleton & 
\textit{Capture}: Distance between landmarks on bodies \par{}\vspace{3pt}
\textit{Inference}: Body segments
\\
\midrule
\textbf{Determine how these objects are made legible to the system} & 
System inputs are markers that are placed on body landmarks as specified by a body model. Multiple calibrated cameras sense each marker, which allows the XYZ position of markers to be inferred. Then, the body model is then used to infer the position and motion of the underlying skeleton. &
System inputs are body landmarks identifiable from the surface of the body, which a tape measure is used to measure the distance between. Both the identification of landmarks and the application of tape measures are informed by public health guides to anthropometric measurement.\\
\midrule
\textbf{Examine how ground truth is and has historically been established} & 

\textit{Historically:} The earliest ground truth data on BSPs used in motion capture relied on dismembering dead bodies or implanting pins into the bones of living bodies in an attempt to measure BSPs directly. Since then, researchers have relied primarily on concurrent validity assessments which show that modern mocap systems produce measurements similar to existing systems~\cite{harvey-cadaver}.\par{}\vspace{3pt}
\textit{Current Context}: Our audit followed the ground truthing practices of the mocap practitioners at our field site: ball-and-stick camera calibration, review of marker placement by multiple researchers, and concurrent recording of motion with markers, video cameras, and a depth camera. &

\textit{Historically}: Units of measurement were originally socially constructed (e.g., based on easy-to-measure and relevant lengths, like armspan) and then standardized (e.g., by measuring how far light travels in a specific fraction of a second). Anthropometric measurement landmarks were originally chosen to study variation in physical anthropology, and later, to aid biometric identification in criminal justice and for eugenic purposes~\cite{maguire2009birth}. Current body landmarks are used to produce measurements of physical health and early childhood development~\citepalias{HHS}.\looseness=-1 
\par{}\vspace{3pt}
\textit{Current Context}: Our audit followed anthropometric guides produced by the U.S. government~\citepalias{HHS} as well as analogous ground-truthing practices to those used for mocap, including review of body landmark identification by multiple researchers and concurrent validity assessments of measurement instruments (i.e., comparing tape measures to one another, measuring a known weight on the scale).\\
\midrule
\textbf{Determine what assumptions ``ground truths'' encode} & People have standard body landmarks (corresponding to those present on the bodies of primarily male research subjects) and have a standard amount of soft tissue between markers and the surface of bodies~\cite{keller2023skin}. & 
People have standard body landmarks and body landmarks are equally identifiable on people regardless of their body shape or size.\\
\midrule
\textbf{Consider how these assumptions may produce harms when operationalized in the relevant context of use} & 
We expect to see less fidelity of representation for people who are not male or not thin. This may cause a person to see their body (or a body like theirs) misrepresented or not represented in an entertainment setting. Measurement processes initially used for entertainment may be adapted for higher-stakes settings, and the harms can travel with them. & 
We expect to see less fidelity of representation for people who are not close to the average used to construct anthropometric measurement procedures and norms. Measurements indicating that individuals are further from the average may be used to reify or reinforce eugenicist harms.\\
\bottomrule
\end{tabular}
\caption{The \textbf{Contextual Auditing Matrix}, completed for the context of mocap for creative design and considering both a marker-based motion capture system and a tape measure.}
\label{tab:mocap-matrix}
\end{table*}

\paragraph{Context Reveals Assumptions.} In our field work, we observed that practitioners considered valid representations of bodies to be those that were faithful to the proportions and movements of subjects on the mocap stage. To produce measurements seen as valid, practitioners attended to the calibration of cameras, the proper placement of markers on subjects' bodies, the management of reflective objects on the stage that might interfere with mocap recordings, and areas of the mocap stage with greater and lesser accuracy and reliability (due to camera placement). We observed that, while practitioners undertook practices to produce what they would see as valid measurements, they ultimately took validity as an article of faith. They judged when a set of measurements ``look[ed] right''  for assurances that their systems were working properly. They looked for areas of the mocap stage where measurements ``g[o]t wonky'' to decide when to recalibrate. By ``wonky,'' practitioners meant that the dimensions and positions of skeletal elements fell out of proportion in the mocap software---elongated necks, twisted limbs, and foreshortened arms were clues practitioners used to identify a need to recalibrate the system. However, in the absence of egregious distortions, they trusted that the calibrated mocap system produced accurate measurements rather than verifying those measurements themselves (by, for example, measuring subjects with an alternate apparatus).\looseness=-1

Our field work thus directed us towards key assumptions encoded in the practice of mocap for creative design and disclosed how ground truth was established; namely that placing markers consistently and calibrating the mocap stage correctly produces valid measurements for any human subject, and that, when they occur, invalid measurements can be identified heuristically. To interrogate these assumptions, we chose to directly audit a basic set of measurements produced by the mocap system: BSPs (specifically, segment length).\looseness=-1

\paragraph{Assumptions Reveal Facets.}
BSPs are an appropriate basis for a contextual audit of a mocap system because it is easy to reason about what they should look like: the BSPs of an individual should be stable over a short period of time, and estimates of BSPs produced by a mocap system should be similar to estimates of BSPs produced by alternative measurement modalities. Based on the OptiTrack documentation, analyzed through the lens of our previously conducted literature review~\cite{harvey-cadaver}, we identified several facets across which to audit the mocap system. The system should perform equally well (e.g., produce equally stable measurements) across the facets of \textit{body size}, \textit{sex}, and \textit{time}.\looseness=-1

We chose these facets based on the assumptions encoded in marker-based mocap systems. As \citet{harvey-cadaver} outline, and as our review of the OptiTrack documentation confirms, mocap systems were built using data from a small number of bodies that were overwhelmingly male and thin. This means that different types of bodies may be more likely to be misrepresented by mocap systems. We chose to focus on the facets of body size and sex in particular because the mocap system we worked with was a marker-based system, and body size and sex are both likely to affect the location of body landmarks (and, correspondingly, marker placement). We also chose to include time as a facet in order to capture variation related to the practice of body measurement using mocap (e.g., room calibration or marker placement on the human subject by experimenters) as well as potential variation in the human subject (e.g., weight gain, change in posture due to injury).\looseness=-1

\subsection{Incorporating Symmetry}\label{subsec:case-study:symmetry}
There is no single agreed-upon system of ``ground truthing'' BSPs. For example, consider the distance between a person's shoulders. One could measure the length of the collarbone, or the straight-line distance between shoulder blades. One could determine the length based only on bones (which are not directly observable while the bones are contained within the body), or one could take soft tissue into account. Universal ground truth values for BSPs are unknown---and potentially unknowable. Thus, we did not have access to a definitive ground truth against which to compare measurements produced by the mocap system. Instead, following practices common in concurrent validity assessments, we compared BSPs produced by the mocap system to anthropometric measurements produced using a tape measure. We chose this alternative modality because it is free of many of the starting assumptions of mocap technology and was easily accessible to us (unlike, e.g., an X-ray machine). Our approach differed from a typical concurrent validity assessment because we incorporated the concept of symmetry. That is, we built upon the understanding that neither the tape measure nor the mocap system could or should be taken as a ground truth. Rather than interrogate the OptiTrack mocap system by taking tape measured BSPs as a ground truth, we used the framework of contextual auditing to examine the assumptions underlying \textit{both} systems. This analysis is summarized in Table~\ref{tab:mocap-matrix}. Then, in our empirical analysis, we built on the notion of symmetry to compare how the different measurements varied with respect to one another, but did not characterize this as systematic bias from a ground truth. In this way, incorporating the concept of symmetry allowed us to conduct a contextual AI audit even in the absence of an observable ground truth.\looseness=-1

\subsection{Audit Protocol} 
After developing an understanding of the context and assumptions surrounding mocap for creative design, we constructed an audit protocol to test the stability of BSPs across our identified facets. The experimental protocol consisted of two measurement phases. In Phase 1, participants' measurements were recorded using traditional anthropometric techniques and equipment~\citepalias{HHS}. In Phase 2, participants' measurements were recorded using the mocap system. To conduct a contextually grounded audit, we embedded ourselves within the mocap environment, strictly adhering to the laboratory's local practices. While the fundamental procedures align with the official OptiTrack Motive documentation,\footnote{\url{https://docs.optitrack.com/motive/skeleton-tracking}} the specific workflows were executed according to internal facility protocols to ensure consistency with the site's standard operating procedures.\looseness=-1

The protocol begins with participants' arrival at the field site, at which time they were asked to provide informed consent to participate, given the opportunity to ask questions about the study, and assigned a participant ID to separate the collected data from the personal information used to manage their enrollment. Participants were then issued a mocap suit---a form-fitting spandex bodysuit with a Velcro exterior appropriate for affixing the markers used in mocap. The mocap suit also facilitated the collection of body measurements in Phase 1, as it removed any obstructions that might have been introduced by structured or bulky clothing.\looseness=-1

\paragraph{Tape Measure.} For Phase 1, each set of body measurements was taken by two members of the research team. In each pair, one researcher took the role of primary measurer, while the other recorded measurements and assisted in measuring wingspan (a two-person job). All measurements except for weight were read aloud by the primary measurer and then verbally confirmed and recorded by the assistant measurer. In most, but not all, cases, we matched the sex of the primary measurer to the sex of the participant that they were measuring. Body measurement protocols were drawn from the National Health and Nutrition Examination Survey~\citepalias{HHS}. All measurements were taken using a soft tape measure, except for weight, which was measured with a digital scale. The specific measurements recorded, listed in the order in which they were taken, were: standing height, wingspan, sitting height, upper leg length, biacromial breadth, upper arm length, upper arm circumference, abdominal circumference, hip circumference, thigh circumference, head circumference, and weight. The full measurement protocol is available in Appendix A.

\paragraph{Mocap System.} One researcher (the technical lead) performed daily room calibration in accordance with standard procedures. This involved clearing the mocap stage of reflective noise and performing a wand wave calibration using the OptiTrack large space calibration wand (a rigid pole with markers affixed to known locations). The wand was moved in a figure-8 pattern until a minimum of 10,000 samples were registered per camera. The team adhered to a strict error threshold, proceeding only when the system reported an ``Exceptional'' rating (approximate mean error $\leq$ 0.6 mm). The ground plane was defined using an L-shaped calibration square aligned with the participants' walking direction.\looseness=-1

Participants were fitted with a standard mocap suit (cap, gloves, and shoes) secured with Velcro. The research team attached 41 retroreflective markers to the suit following the standard OptiTrack ``Baseline 41'' body model. One of the mocap practitioners at the facility, a specialist with extensive experience in the entertainment industry, performed periodic quality control checks to ensure marker placement aligned with industry standards. Following marker placement, the technical lead conducted subject calibration. Participants assumed a static T-pose (standing with feet shoulder-width apart, arms raised to 90 degrees) to allow the Motive software to infer the body model automatically. Neither manual alignment of the skeleton nor dynamic calibration was performed. The technical lead then assigned a generic avatar for real-time visualization purposes.\looseness=-1

Once calibrated, participants followed a scripted protocol of distinct everyday movements: performing a T-pose, walking, sitting, rotating, squatting, swinging arms, making circles with arms, jumping, kicking, carrying a 10-lb kettlebell (in front and to the side of the body), carrying a large exercise ball, stepping onto a platform, bouncing a kickball, cross toe touching, running in place, using a cell phone, pointing, checking the time on a watch, rubbing hands together, crossing arms, and planking. These movements were chosen because they represent everyday body movements that may feature in entertainment products using mocap, as well as actions that might induce errors in mocap systems (as they may cause markers to be occluded or change a participant's center of gravity). While the technical lead operated the Motive software and recording from the control desk, two additional researchers read out the movement protocol and demonstrated movements where appropriate. We recorded the mocap data produced via the OptiTrack system as well as auxiliary data; namely, a depth camera recording of the mocap stage (OAK-D) and a screen capture of the live Motive interface. To export BSPs from the system, we exported the Motive recording as a CSV containing, for every frame, the 3D position and orientation quaternion of each skeletal element produced by the Baseline 41 body model. From these files we read the endpoints of each BSP as the Euclidean distance between two endpoints: shoulder width as left to right upper arm; upper arm length as upper arm to forearm; lower arm length as forearm to hand; upper leg length as thigh to shin; lower leg length as shin to foot; and standing height as the mean of the head-to-foot distances on the left and right sides. Distances were computed on the first 10 frames of each take, captured while the participant held the static T-pose used for subject calibration, and averaged to yield a single BSP value per measurement per session. The extraction script is available in our public repository\footnote{\url{https://github.com/HaukeCornell/Audit_Skeletons}} and the full mocap protocol is available in Appendix B.\looseness=-1

\subsection{Data Analysis}
Our analysis focuses on a set of BSPs: the estimates of standing and sitting height, wingspan, biacromial breadth (distance between shoulders) and upper arm and leg length produced by OptiTrack and by tape measurements. One intuitively desirable characteristic of BSP measurement is that it should be reliable. In other words, the BSPs of an individual should be stable over a short period of time or when produced by different measurement instruments. Therefore, we focus on the difference between measurements---e.g., across measurement techniques or over time---as the dependent variable in our analysis. The independent variables are the facets of body size (operationalized as BMI, weight and body segment size),\footnote{Body Mass Index, or BMI, is a ratio of weight to height ($\frac{\text{weight (kg)}}{\text{height (m)}^2}$). BMI was originally developed for white male bodies and intended as a population-level statistic~\cite{Pray2023-ml}. Today, BMI is applied to bodies of all types as a measure of individual health. Although BMI is problematic, it is correlated with the presence of soft tissue (i.e., skin, fat, and muscle), and soft tissue is associated with mocap error~\cite{Jeong2023}. BMI is less invasive than other approaches for measuring soft tissue, such as skinfold calipers. We supplement BMI with weight and body segment size in order to avoid relying on BMI alone.\looseness=-1} sex (self-reported), and time (first or second session), as well as measurement modality (mocap vs. tape measure). Within each session, we compare measurements of the same body part across the different modalities. We also compare measurements of the same body part across different sessions. We systematically compare how reliable our measurements are with respect to facets of interest. Noting that our results are fundamentally exploratory, not confirmatory, we 
test the following hypotheses:\looseness=-1
\begin{itemize}
\item[\textbf{H0}] Mocap measurements for the same participant are reliable across time points. 
\item[\textbf{H1}] The relationship between tape measured and mocap measurements is reliable across participant body size.
\item[\textbf{H2}] The relationship between tape measured and mocap measurements is reliable across participant sex. 
\item[\textbf{H3}] The relationship between tape measured and mocap measurements is reliable across time points.  
\end{itemize}

\subsubsection{Statistical Models.}
We build upon a statistical model commonly used in mocap studies and reliability engineering more broadly: the Bland-Altman limits of agreement (LOA)~\cite{bland_altman}. We extend the Bland-Altman LOA with nested regression models to explore the relationship between reliability and facets of interest. Calculating the Bland-Altman LOA involves regressing the difference between paired measurements on the average of the paired measurement. We focus specifically on comparisons of measurements taken of the same body part of the same subject. To determine whether a measurement should be considered reliable across a modality or facet of interest, we define reliability as having regression coefficients that are not statistically significantly different from zero ($p \geq 0.05$, corrected for multiple comparisons). If a measurement is reliable, measurement differences do not vary with respect to the included features.\looseness=-1

Across-session, we compare across measurements produced by modality $x$ for individual $i$ and body part $j$ in Sessions 1 and 2. We apply the following model, iteratively adding explanatory variables: 
\begin{small}
$$x_{ij1} - x_{ij2} \sim  \frac{x_{ij1} + x_{ij2}}{2} +  \text{BMI}_{it} + \text{weight}_{it} + \text{sex}_{it}$$
\end{small}
Within-session, we compare across measurement modality $x$ and $y$, for individual $i$ and body part $j$ in session $t$. We apply the following model, again iteratively adding explanatory variables: 
\begin{small}
$$x_{ijt} - y_{ijt} \sim \frac{ x_{ijt} + y_{ijt}}{2} + \text{session}_{t} + \text{BMI}_{it} +  \text{weight}_{it} + \text{sex}_{it}$$
\end{small}
Following typical practice for Bland-Altman LOA, a nonzero coefficient for the first term would indicate systematic variation across modalities with respect to the target of measurement. The other variables (session, weight, sex, etc.) would capture systematic variation in measurement reliability across methods with respect to those facets, all else constant. We pre-registered our plan prior to conducting data analysis.\footnote{\url{https://bit.ly/osfxy4jg}} Our analysis code is described and linked in Appendix C.\looseness=-1

\subsection{Audit Results}
We applied the protocol to 24 unique participants, 12 of whom were measured during two sessions approximately one month apart. The participants ranged in age (21--63, median: 32), height (153--193 cm, median: 171 cm), and weight (45.5--115.5 kg, median: 71 kg), and sex (male: 11, female: 13). Because of the large number of variables and consistent findings, we briefly summarize our key results here. Our full results are available in Appendix D.\looseness=-1

\textit{\textbf{H0}: Mocap measurements for the same participant are reliable across time points.} This mostly holds, except for wingspan and biacromial breadth, where the difference increases as the average measurement increases.

\textit{\textbf{H1}: The relationship between tape measured and mocap measurements is reliable across participant body size.} This mostly holds, with two exceptions. First, the difference between biacromial breadth as measured by a tape measure vs. OptiTrack increases as average biacromial breadth decreases. This holds whether or not time, BMI, weight, or sex are accounted for. Second, for standing height, the difference increases as the average measurement increases. Conditional on average standing height, this deviation increases as BMI increases. Conditional on those, it decreases as weight increases. This holds whether or not sex is accounted for.\looseness=-1

\textit{\textbf{H2:} The relationship between tape measured and mocap measurements is reliable across participant sex.} This holds (i.e., no statistically significant differences).

\textit{\textbf{H3:} The relationship between tape measured and mocap measurements is reliable across time points.} This holds.

We emphasize that our results are fundamentally \textit{exploratory} and should only be interpreted as such. Any detected systematic differences across facets should be seen as hypothesis-generating for future analysis, not evidence of explicit bias of either the OptiTrack mocap system or of a tape measure.\looseness=-1

\section{Discussion}\label{sec:5-discussion}
The contextual audit framework allows auditors to explore measurements in the context of the practices that produced them. In this way, contextual audits provide a way of answering the two fundamental questions raised by AI auditing: ``what inputs are relevant for identifying patterns in a system’s actual outputs?'' and ``what should a system’s nominal outputs look
like?'' In our case study audit (\S\ref{sec:4-case-study}), the contextual framework enabled us to identify assumptions underlying BSP measurement as conducted by mocap systems as well as facets across which to audit. By incorporating the concept of symmetry from STS, we were able to conduct a contextual audit of mocap's ability to measure BSPs, despite the fact that there is no single agreed-upon approach for ``ground-truthing'' BSPs---that is, in the context of BSPs, ground truth can be constructed or approximated in multiple different ways. We did this by using an alternate measurement modality to serve as a provisional ground truth and treating that provisional ground truth as the nominal outputs according to which we evaluated actual system outputs. In turn, that modality could also be evaluated using mocap: under the principle of symmetry, each measurement modality can serve as ground truth for the other. 

The value of symmetry in auditing skeleton inference is illustrated by \textbf{H1} above, where the actual and nominal outcomes vary in two cases. First, the variation between mocap and tape measured estimates of biacromial breadth increases as segment length decreases. Applying the principle of symmetry, we can explain this variation from the position that either modality is potentially true while the other is not. The tape measurements may be particularly sensitive to subjects' posture when measuring across their shoulders, \textit{or} there may be systematic irregularities having to do with shoulder width in the mocap system.\footnote{Or tape measurements may be applied inconsistently across researchers, or some combination of these and other processes.} Second, the variation between mocap and tape measured estimates of height increases as height increases. There could be systematic error in the anthropometric protocol used to measure standing height---perhaps the protocols are poorly calibrated in ways that are only apparent when measuring larger values (standing height being the largest set of values measured with the tape measure). Symmetrically, there could be systematic error in the mocap system, as suggested by \citet{harvey-cadaver}, for taller individuals with higher BMIs, who were under-represented in the original BSP calculations. The audit does not resolve the question of what causes these unreliable measurements. Rather, applying symmetry demonstrates how audit results can still be used to surface practical next steps for the audit process when ground truth is unknown, unknowable, or contested.\looseness=-1

\paragraph{Audit Implications.}
As discussed in \S\ref{subsec:background:ai-auditing}, a primary goal of AI auditing is accountability~\cite{birhane_ai_2024, bovens}. Thus, we make our expectations of accountability explicit here. As an exploratory audit, we do not directly identify disparities that mocap providers must ``fix'' (although future confirmatory audits may do so). Instead, we call on both mocap providers and researchers to consider the assumptions underlying mocap systems in their work; to test system performance across the relevant facets of sex and body size; and to go beyond straightforward concurrent validity assessments in their analyses. More broadly, we call on researchers and practitioners to consider ecological validity by using qualitative methods to understand the context in which systems are actually designed and used before conducting AI audits.\looseness=-1

\paragraph{Future Work.} We identify several opportunities for future work. First, we recognize the importance of building on our case study audit with additional studies designed to resolve the questions raised by our symmetrical approach. Second, we call for contextual audits to be adopted as an approach for auditing other AI systems. We argue that it is particularly urgent to adopt a contextual audit approach for settings where measurement practices carry heavy traditions of dehumanizing, pseudoscientific, and otherwise harmful assumptions (e.g., in hiring, where AI/ML-based tools have been shown to have eugenicist and pseudoscientific roots~\cite{sloane_silicon_2022}). AI audits that fail to adopt a contextual approach (e.g., hiring audits that do not assess tools in their actual contexts of use) may instead perpetuate the harms they intend to mitigate.\looseness=-1

While we emphasize that our case study audit is exploratory, we can still identify opportunities for future audits into mocap systems, especially in higher-stakes settings. We suggest that AI audits should focus on facets that displayed lower stability in our case study audit by researching how well mocap systems perform for bodies with narrower shoulders as well as for larger bodies. This is particularly important in the context of prior work by \citet{harvey-cadaver}, who found that early research into mocap systems, which has stabilized into current practices, overwhelmingly included thin bodies---for example, just 5\% of subjects in early research on soft tissue artifacts, which occur when skin, muscle, and fat cause markers on the surface of subjects' bodies to move independently of subjects' bones, were noted as ``overweight'' (according to BMI). Modern mocap research acknowledges, but still does not address, that mocap methods are often invalid for subjects with more soft tissue~\cite{chu_mocap}; it may attribute errors in mocap systems ``to the anatomy of the [subject]'' and not necessarily to the mocap system itself~\cite{paloschi_mocap}. Thus, future mocap audits with a continued focus on body size as a facet are critical.\looseness=-1

Finally, we call for additional research into how to conduct audits when ground truth is unknown, unknowable, or contested. While the concept of symmetry can help identify areas of instability for future audits, as we show in our study, confirmatory audits may still require a more explicit ground truth.\looseness=-1

\paragraph{Limitations.} 
The case study audit we describe in \S\ref{sec:4-case-study} is intended to illustrate our conceptual contribution: a rigorous methodology to incorporate the social practice focus of contextual audits into the statistical framework of AI auditing. The results we share are fundamentally exploratory and should not be taken as a definitive endorsement or critique of either OptiTrack or tape measures.\looseness=-1

Our experimental setup highlights many of the idiosyncratic choices that become encoded in typical AI audits. The mocap setting specifically requires choices about types of populations (demographics, ranges of body types, ranges of mobility and mobility aids, etc.) that are already poorly cataloged in existing mocap and computer vision development and validation practices~\cite{sloane2025making}. Our study, for example, did not include participants who used mobility aids and primarily (although not deliberately) included participants who fell within the so-called `normal' BMI category. In addition, our small sample size, non-expert measurers, and multiple regressions limit the empirical claims we can make with our data.\looseness=-1

Perhaps most importantly, we note that we are researchers, and not truly real-world system users, i.e., expert practitioners with years of experience in mocap for creative design. As \citet{harvey-cadaver} describe, errors in measurements produced by mocap systems have several sources: while some may be related to the mocap system itself and the assumptions it makes, others are attributable to the operators of a mocap system. In particular, operators produce \textit{marker placement errors}, which occur when the markers used by a mocap system are not placed correctly on the body landmark specified by a body model~\cite{harvey-cadaver}. We anticipated that the magnitude of marker placement error produced by us, as non-expert mocap users, would be larger than the magnitude of marker placement error produced by the expert practitioners whose practices we sought to replicate. This complicates how audit results can be interpreted: rather than an interrogation of the system alone, our audit produces an interrogation of the system and its operators (us). We believe that this is likely to be broadly true for audit studies in which researchers assess a system typically used by experts, and call for contextual audits to explicitly include a \textit{reflexive} step in which auditors consider how to disentangle what audit results say about AI systems from what audit results say about auditors.\looseness=-1

\section{Conclusion}\label{conclusion}
We introduce \textbf{contextual audits} as a method for auditing measurements within the context of the practices that produce them. We show how incorporating the concept of \textbf{symmetry} from STS can enable auditors to be explicit about what serves as a ground truth and, if needed, conduct audits when ground truth is unknown, unknowable, or contested. Finally, to demonstrate how contextual and symmetric audits can be conducted in practice, we present a case study of skeleton inference in motion capture.\looseness=-1

\section*{Ethical Statement}
\paragraph{Researcher Positionality.}
We are a multidisciplinary team of researchers, with academic backgrounds in computer science, data science, information science, anthropology, and sociology. Our backgrounds have shaped the disciplinary lenses that we applied to our audit, leading us to draw on the concepts of social practice theory from sociology and reflexivity and symmetry from STS. Our backgrounds have also shaped the questions we asked in our research. While we study mocap systems, no member of the research team has professional experience as a mocap practitioner. Several members of the team, however, have experience in the arts or in creative design. Our bodies, in some ways, look like those that mocap systems were originally designed for (we are all white and cisgender); in other ways, they do not (more than half of us are women). We thus approached our audit as outsiders to mocap for creative design, but with the lived experience of seeing bodies that looked like ours represented (and sometimes misrepresented) in mocap-enabled visual arts.

\paragraph{Ethical Considerations.}
The primary ethical considerations we made for this study were related to the safety and privacy of audit participants. Our study was approved by a university Institutional Review Board. All participants provided written informed consent prior to engaging in the study. Participants were compensated for their time with a \$15 gift card per session attended, an amount set to ensure participants would not feel compelled to participate due to financial need. At the conclusion of the session, all participants were given the opportunity to participate in a semi-structured ethnographic interview regarding their experience of being measured and digitized. Finally, in order to protect participant privacy, all study data was stored separately from participant names using anonymized ID variables and all video data was de-identified using facial blurring. An additional ethical concern relates to the historical uses of anthropometry for race science and eugenics~\cite{clever2023biometry, turda2010race}. Our implementation of anthropometric techniques was strictly limited to vetted, contemporary methods widely deployed in healthcare~\citepalias{HHS}.

\paragraph{Adverse Impact.}
We believe that our study poses only minimal risks to participants. Participants were asked only to conduct everyday movements, such as stepping and bouncing a ball, and were explicitly instructed not to complete movements that felt uncomfortable to them. To mitigate potential emotional harms associated with being weighed, we gave all participants the option of being weighed while facing away from the scale display, and did not read participant weights aloud. Participants also faced the risk of emotional harm from seeing their bodies misrepresented in the mocap software; however, based on exit interviews, we believe that participants did not experience this as a harm.\looseness=-1

Like any audit of an AI system, our study runs the broader risk of inadvertently contributing to improving or legitimizing technology that enables large-scale surveillance or other negative societal effects. We believe that in our case, this risk is minimal, as the audit we present is intended to be an illustrative case study of contextual auditing as opposed to a rigorous audit of the OptiTrack mocap system.\looseness=-1

\section*{Acknowledgments}
This work was supported by funding from the Notre Dame-IBM Technology Ethics Lab. We thank Todd Bryant, Luke DuBois, Kaustav Sarkar, and Harsh Palan for their valuable support.

\bibliography{references}

@article{bovens,
author = {Bovens, Mark},
title = {Analysing and Assessing Accountability: A Conceptual Framework},
journal = {European Law Journal},
volume = {13},
number = {4},
pages = {447-468},
doi = {https://doi.org/10.1111/j.1468-0386.2007.00378.x},
url = {https://onlinelibrary.wiley.com/doi/abs/10.1111/j.1468-0386.2007.00378.x},
eprint = {https://onlinelibrary.wiley.com/doi/pdf/10.1111/j.1468-0386.2007.00378.x},
year = {2007}
}

@inproceedings{mei_addressing_2026,
author = {Mei, Katelyn X. and Choi, Anna Seo Gyeong and Schellmann, Hilke and Sloane, Mona and Koenecke, Allison},
title = {Addressing Auditing Pitfalls in Automatic Speech Recognition Technologies: A Case Study of People with Aphasia},
year = {2026},
isbn = {9798400725968},
publisher = {Association for Computing Machinery},
address = {New York, NY, USA},
url = {https://doi.org/10.1145/3805689.3812320},
doi = {10.1145/3805689.3812320},
booktitle = {Proceedings of the 2026 ACM Conference on Fairness, Accountability, and Transparency},
pages = {3422–3465},
numpages = {44},
location = {
},
series = {FAccT '26}
}

@inproceedings{pruss_ghosting_2023,
	address = {New York, NY, USA},
	series = {{FAccT}},
	title = {Ghosting the {Machine}: {Judicial} {Resistance} to a {Recidivism} {Risk} {Assessment} {Instrument}},
	isbn = {979-8-4007-0192-4},
	shorttitle = {Ghosting the {Machine}},
	url = {https://dl.acm.org/doi/10.1145/3593013.3593999},
	doi = {10.1145/3593013.3593999},
	urldate = {2023-06-13},
	booktitle = {Proceedings of the 2023 {ACM} {Conference} on {Fairness}, {Accountability}, and {Transparency}},
	publisher = {Association for Computing Machinery},
	author = {Pruss, Dasha},
	month = jun,
	year = {2023},
	pages = {312--323},
}

@misc{gosciak2026llmssocialservicesdoes,
      title={LLMs in social services: How does chatbot accuracy affect human accuracy?}, 
      author={Jennah Gosciak and Eric Giannella and Zhaowen Guo and Michael Chen and Allison Koenecke},
      year={2026},
      eprint={2603.11213},
      archivePrefix={arXiv},
      primaryClass={cs.HC},
      url={https://arxiv.org/abs/2603.11213}, 
}

@inproceedings{cheng_how_2022,
	address = {New Orleans LA USA},
	series = {{CHI}},
	title = {How {Child} {Welfare} {Workers} {Reduce} {Racial} {Disparities} in {Algorithmic} {Decisions}},
	isbn = {978-1-4503-9157-3},
	url = {https://dl.acm.org/doi/10.1145/3491102.3501831},
	doi = {10.1145/3491102.3501831},
	language = {en},
	urldate = {2023-07-07},
	booktitle = {{CHI} {Conference} on {Human} {Factors} in {Computing} {Systems}},
	publisher = {ACM},
	author = {Cheng, Hao-Fei and Stapleton, Logan and Kawakami, Anna and Sivaraman, Venkatesh and Cheng, Yanghuidi and Qing, Diana and Perer, Adam and Holstein, Kenneth and Wu, Zhiwei Steven and Zhu, Haiyi},
	month = apr,
	year = {2022},
	pages = {1--22},
}

@article{Kihlstrom_ecological_2021,
author = {John F. Kihlstrom},
title ={Ecological Validity and “Ecological Validity”},
journal = {Perspectives on Psychological Science},
volume = {16},
number = {2},
pages = {466-471},
year = {2021},
doi = {10.1177/1745691620966791},
    note ={PMID: 33593121},
URL = {https://doi.org/10.1177/1745691620966791},
}

@article{feng_has_2022,
	series = {{AAAI}},
	title = {Has {CEO} {Gender} {Bias} {Really} {Been} {Fixed}? {Adversarial} {Attacking} and {Improving} {Gender} {Fairness} in {Image} {Search}},
	volume = {36},
	copyright = {Copyright (c) 2022 Association for the Advancement of Artificial Intelligence},
	issn = {2374-3468},
	shorttitle = {Has {CEO} {Gender} {Bias} {Really} {Been} {Fixed}?},
	url = {https://ojs.aaai.org/index.php/AAAI/article/view/21445},
	doi = {10.1609/aaai.v36i11.21445},
	language = {en},
	number = {11},
	urldate = {2023-04-07},
	journal = {Proceedings of the AAAI Conference on Artificial Intelligence},
	author = {Feng, Yunhe and Shah, Chirag},
	month = jun,
	year = {2022},
	pages = {11882--11890},
}

@article{sweeney_discrimination_2013,
	title = {Discrimination in online ad delivery},
	volume = {56},
	issn = {0001-0782, 1557-7317},
	url = {https://dl.acm.org/doi/10.1145/2447976.2447990},
	doi = {10.1145/2447976.2447990},
	language = {en},
	number = {5},
	urldate = {2023-10-25},
	journal = {Communications of the ACM},
	author = {Sweeney, Latanya},
	month = may,
	year = {2013},
	pages = {44--54},
}

@misc{compas,
url={https://www.propublica.org/article/machine-bias-risk-assessments-in-criminal-sentencing},
title={Machine Bias},
author={Julia Angwin and Jeff Larson and Surya Mattu and Lauren Kirchner},
year=2016,
howpublished={ProPublica}}

@inproceedings{vecchione_algorithmic_2021,
	address = {-- NY USA},
	series = {{EAAMO}},
	title = {Algorithmic {Auditing} and {Social} {Justice}: {Lessons} from the {History} of {Audit} {Studies}},
	isbn = {978-1-4503-8553-4},
	shorttitle = {Algorithmic {Auditing} and {Social} {Justice}},
	url = {https://dl.acm.org/doi/10.1145/3465416.3483294},
	doi = {10.1145/3465416.3483294},
	language = {en},
	urldate = {2023-10-25},
	booktitle = {Equity and {Access} in {Algorithms}, {Mechanisms}, and {Optimization}},
	publisher = {ACM},
	author = {Vecchione, Briana and Levy, Karen and Barocas, Solon},
	month = oct,
	year = {2021},
	pages = {1--9},
}

@Article{wang_inadequacy_2025,
author={Wang, Angelina
and Ho, Daniel E.
and Koyejo, Sanmi},
title={The inadequacy of offline large language model evaluations: A need to account for personalization in model behavior},
journal={Patterns},
year={2025},
month={Dec},
day={12},
publisher={Elsevier},
volume={6},
number={12},
issn={2666-3899},
doi={10.1016/j.patter.2025.101397},
url={https://doi.org/10.1016/j.patter.2025.101397}
}

@inproceedings{harvey_framework_2025,
	address = {Athens Greece},
	series = {{FAccT}},
	title = {A {Framework} for {Auditing} {Chatbots} for {Dialect}-{Based} {Quality}-of-{Service} {Harms}},
	isbn = {979-8-4007-1482-5},
	url = {https://dl.acm.org/doi/10.1145/3715275.3732137},
	doi = {10.1145/3715275.3732137},
	language = {en},
	urldate = {2025-07-01},
	booktitle = {Proceedings of the 2025 {ACM} {Conference} on {Fairness}, {Accountability}, and {Transparency}},
	publisher = {ACM},
	author = {Harvey, Emma and Kizilcec, Rene F. and Koenecke, Allison},
	month = jun,
	year = {2025},
	pages = {2025--2039},
}

@inproceedings{fleisig_linguistic_2024,
	address = {Miami, Florida, USA},
	series = {{EMNLP}},
	title = {Linguistic {Bias} in {ChatGPT}: {Language} {Models} {Reinforce} {Dialect} {Discrimination}},
	shorttitle = {Linguistic {Bias} in {ChatGPT}},
	url = {https://aclanthology.org/2024.emnlp-main.750},
	doi = {10.18653/v1/2024.emnlp-main.750},
	language = {en},
	urldate = {2025-03-28},
	booktitle = {Proceedings of the 2024 {Conference} on {Empirical} {Methods} in {Natural} {Language} {Processing}},
	publisher = {Association for Computational Linguistics},
	author = {Fleisig, Eve and Smith, Genevieve and Bossi, Madeline and Rustagi, Ishita and Yin, Xavier and Klein, Dan},
	month = nov,
	year = {2024},
	pages = {13541--13564},
}

@article{hofmann_ai_2024,
	series = {Nature},
	title = {{AI} generates covertly racist decisions about people based on their dialect},
	volume = {633},
	issn = {0028-0836, 1476-4687},
	url = {https://www.nature.com/articles/s41586-024-07856-5},
	doi = {10.1038/s41586-024-07856-5},
	language = {en},
	number = {8028},
	urldate = {2025-04-01},
	journal = {Nature},
	author = {Hofmann, Valentin and Kalluri, Pratyusha Ria and Jurafsky, Dan and King, Sharese},
	month = sep,
	year = {2024},
	pages = {147--154},
}

@inproceedings{wilson_gender_2024,
	series = {{AIES}},
	title = {Gender, {Race}, and {Intersectional} {Bias} in {Resume} {Screening} via {Language} {Model} {Retrieval}},
	url = {https://ojs.aaai.org/index.php/AIES/article/view/31748},
	language = {en},
	booktitle = {Proceedings of the {Seventh} {AAAI}/{ACM} {Conference} on {AI}, {Ethics}, and {Society} ({AIES2024})},
	author = {Wilson, Kyra and Caliskan, Aylin},
	month = oct,
	year = {2024},
}

@inproceedings{imana_external_2025,
	address = {Athens Greece},
	series = {{FAccT}},
	title = {External {Evaluation} of {Discrimination} {Mitigation} {Efforts} in {Meta}'s {Ad} {Delivery}},
	isbn = {979-8-4007-1482-5},
	url = {https://dl.acm.org/doi/10.1145/3715275.3732170},
	doi = {10.1145/3715275.3732170},
	language = {en},
	urldate = {2025-07-01},
	booktitle = {Proceedings of the 2025 {ACM} {Conference} on {Fairness}, {Accountability}, and {Transparency}},
	publisher = {ACM},
	author = {Imana, Basileal and Shen, Zeyu and Heidemann, John and Korolova, Aleksandra},
	month = jun,
	year = {2025},
	pages = {2616--2629},
}

@misc{pr_xsens,
title={Motion Capture and the Future of Occupational Health and Safety},
url={https://www.xsens.com/resources/blog/motion-capture-and-the-future-of-occupational-health-and-safety}, 
author={Xsens},
year=2024}

@inproceedings{sandvig_auditing_2014,
	address = {Seattle, WA},
	title = {Auditing {Algorithms}: {Research} {Methods} for {Detecting} {Discrimination} on {Internet} {Platforms}},
	shorttitle = {Auditing {Algorithms}},
	url = {https://www.semanticscholar.org/paper/Auditing-Algorithms-%3A-Research-Methods-for-on-Sandvig-Hamilton/b7227cbd34766655dea10d0437ab10df3a127396},
	urldate = {2023-04-06},
	booktitle = {"{Data} and {Discrimination}: {Converting} {Critical} {Concerns} into {Productive} {Inquiry},” a preconference at the 64th {Annual} {Meeting} of the {International} {Communication} {Association}},
	author = {Sandvig, Christian and Hamilton, Kevin and Karahalios, K. and Langbort, Cédric},
	month = may,
	year = {2014},
}

@inproceedings{birhane_ai_2024,
author = { Birhane, Abeba and Steed, Ryan and Ojewale, Victor and Vecchione, Briana and Raji, Inioluwa Deborah },
booktitle = { 2024 IEEE Conference on Secure and Trustworthy Machine Learning (SaTML) },
title = {{ AI auditing: The Broken Bus on the Road to AI Accountability }},
year = {2024},
volume = {},
ISSN = {},
pages = {612-643},
doi = {10.1109/SaTML59370.2024.00037},
url = {https://doi.ieeecomputersociety.org/10.1109/SaTML59370.2024.00037},
publisher = {IEEE Computer Society},
address = {Los Alamitos, CA, USA},
month =apr}

@article{harvey-cadaver,
    author = {Harvey, Emma and Sandhaus, Hauke and Jacobs, Abigail Z.\ and Moss, Emanuel and Sloane, Mona},
    title = {\emph{The Cadaver in the Machine}: The Social Practices of Measurement and Validation in Motion Capture Technology},
    journal = {Proc. ACM CHI},
    year = {2024}
}

@inproceedings{sloane2025making,
  title={Making Bodies: Assumptions in the Design and Validation of Motion Capture Technology},
  author={Sloane, Mona and Jacobs, Abigail Z and Moss, Emanuel},
  booktitle={Proceedings of the AAAI/ACM Conference on AI, Ethics, and Society},
  volume={8},
  pages={2410--2418},
  year={2025}
}

@INPROCEEDINGS{geiger_trace_2011,
  author={Geiger, R. Stuart and Ribes, David},
  booktitle={2011 44th Hawaii International Conference on System Sciences}, 
  title={Trace Ethnography: Following Coordination through Documentary Practices}, 
  year={2011},
  volume={},
  number={},
  pages={1-10},
  doi={10.1109/HICSS.2011.455}}

@inproceedings{becerrasandoval_historical_2025,
author = {Becerra Sandoval, Juana Catalina and Jing, Felicia S.},
title = {Historical Methods for AI Evaluations, Assessments, and Audits},
year = {2025},
isbn = {9798400714825},
publisher = {Association for Computing Machinery},
address = {New York, NY, USA},
url = {https://doi.org/10.1145/3715275.3732093},
doi = {10.1145/3715275.3732093},
booktitle = {Proceedings of the 2025 ACM Conference on Fairness, Accountability, and Transparency},
pages = {1371–1386},
numpages = {16},
location = {
},
series = {FAccT '25}
}

@article{HAN2013131,
title = {A vision-based motion capture and recognition framework for behavior-based safety management},
journal = {Automation in Construction},
volume = {35},
pages = {131-141},
year = {2013},
issn = {0926-5805},
doi = {https://doi.org/10.1016/j.autcon.2013.05.001},
url = {https://www.sciencedirect.com/science/article/pii/S0926580513000514},
author = {SangUk Han and SangHyun Lee}
}

@article{topham_human_2022,
author = {Topham, Luke K. and Khan, Wasiq and Al-Jumeily, Dhiya and Hussain, Abir},
title = {Human Body Pose Estimation for Gait Identification: A Comprehensive Survey of Datasets and Models},
year = {2022},
issue_date = {June 2023},
publisher = {Association for Computing Machinery},
address = {New York, NY, USA},
volume = {55},
number = {6},
issn = {0360-0300},
url = {https://doi.org/10.1145/3533384},
doi = {10.1145/3533384},
journal = {ACM Comput. Surv.},
month = dec,
articleno = {120},
numpages = {42}
}

@article{durkin_analysis_2003,
	title = {Analysis of {Body} {Segment} {Parameter} {Differences} {Between} {Four} {Human} {Populations} and the {Estimation} {Errors} of {Four} {Popular} {Mathematical} {Models}},
	volume = {125},
	issn = {0148-0731, 1528-8951},
	url = {https://asmedigitalcollection.asme.org/biomechanical/article/125/4/515/447637/Analysis-of-Body-Segment-Parameter-Differences},
	doi = {10.1115/1.1590359},
	language = {en},
	number = {4},
	urldate = {2023-05-02},
	journal = {Journal of Biomechanical Engineering},
	author = {Durkin, Jennifer L. and Dowling, James J.},
	month = aug,
	year = {2003},
	pages = {515--522},
}

@inproceedings{buolamwini_gender_2018,
	series = {{FAT}*},
	title = {Gender {Shades}: {Intersectional} {Accuracy} {Disparities} in {Commercial} {Gender} {Classification}},
	shorttitle = {Gender {Shades}},
	url = {https://proceedings.mlr.press/v81/buolamwini18a.html},
	language = {en},
	urldate = {2023-04-06},
	booktitle = {Proceedings of the 1st {Conference} on {Fairness}, {Accountability} and {Transparency}},
	publisher = {PMLR},
	author = {Buolamwini, Joy and Gebru, Timnit},
	month = jan,
	year = {2018},
	pages = {77--91},
}

@article{obermeyer_dissecting_2019,
	title = {Dissecting racial bias in an algorithm used to manage the health of populations},
	volume = {366},
	issn = {0036-8075, 1095-9203},
	url = {https://www.science.org/doi/10.1126/science.aax2342},
	doi = {10.1126/science.aax2342},
	language = {en},
	number = {6464},
	urldate = {2023-04-11},
	journal = {Science},
	author = {Obermeyer, Ziad and Powers, Brian and Vogeli, Christine and Mullainathan, Sendhil},
	month = oct,
	year = {2019},
	pages = {447--453},
}

@article{sloane_silicon_2022,
	title = {A {Silicon} {Valley} love triangle: {Hiring} algorithms, pseudo-science, and the quest for auditability},
	volume = {3},
	issn = {26663899},
	shorttitle = {A {Silicon} {Valley} love triangle},
	url = {https://linkinghub.elsevier.com/retrieve/pii/S2666389921003081},
	doi = {10.1016/j.patter.2021.100425},
	language = {en},
	number = {2},
	urldate = {2022-02-15},
	journal = {Patterns},
	author = {Sloane, Mona and Moss, Emanuel and Chowdhury, Rumman},
	month = feb,
	year = {2022},
	pages = {100425},
}

@book{bandalos2018measurement,
  title={Measurement theory and applications for the social sciences},
  author={Bandalos, Deborah L},
  year={2018},
  publisher={Guilford Publications}
}

@article{bandy_problematic_2021,
	series = {{CSCW}},
	title = {Problematic {Machine} {Behavior}: {A} {Systematic} {Literature} {Review} of {Algorithm} {Audits}},
	volume = {5},
	shorttitle = {Problematic {Machine} {Behavior}},
	url = {https://doi.org/10.1145/3449148},
	doi = {10.1145/3449148},
	number = {CSCW1},
	urldate = {2023-04-07},
	journal = {Proceedings of the ACM on Human-Computer Interaction},
	author = {Bandy, Jack},
	month = apr,
	year = {2021},
	pages = {74:1--74:34},
}

@inproceedings{rismani_plane_2023,
	address = {New York, NY, USA},
	series = {{CHI} '23},
	title = {From {Plane} {Crashes} to {Algorithmic} {Harm}: {Applicability} of {Safety} {Engineering} {Frameworks} for {Responsible} {ML}},
	isbn = {9781450394215},
	shorttitle = {From {Plane} {Crashes} to {Algorithmic} {Harm}},
	url = {https://dl.acm.org/doi/10.1145/3544548.3581407},
	doi = {10.1145/3544548.3581407},
	urldate = {2023-07-11},
	booktitle = {Proceedings of the 2023 {CHI} {Conference} on {Human} {Factors} in {Computing} {Systems}},
	publisher = {Association for Computing Machinery},
	author = {Rismani, Shalaleh and Shelby, Renee and Smart, Andrew and Jatho, Edgar and Kroll, Joshua and Moon, AJung and Rostamzadeh, Negar},
	month = apr,
	year = {2023},
	pages = {1--18},
}

@article{bertrand_emily_2004,
Author = {Bertrand, Marianne and Mullainathan, Sendhil},
Title = {Are Emily and Greg More Employable Than Lakisha and Jamal? A Field Experiment on Labor Market Discrimination},
Journal = {American Economic Review},
Volume = {94},
Number = {4},
Year = {2004},
Month = {September},
Pages = {991–1013},
DOI = {10.1257/0002828042002561},
URL = {https://www.aeaweb.org/articles?id=10.1257/0002828042002561}}

@inproceedings{blodgett_language_2020,
	address = {Online},
	series = {{ACL}},
	title = {Language ({Technology}) is {Power}: {A} {Critical} {Survey} of “{Bias}” in {NLP}},
	shorttitle = {Language ({Technology}) is {Power}},
	url = {https://www.aclweb.org/anthology/2020.acl-main.485},
	doi = {10.18653/v1/2020.acl-main.485},
	language = {en},
	urldate = {2024-03-07},
	booktitle = {Proceedings of the 58th {Annual} {Meeting} of the {Association} for {Computational} {Linguistics}},
	publisher = {Association for Computational Linguistics},
	author = {Blodgett, Su Lin and Barocas, Solon and Daumé Iii, Hal and Wallach, Hanna},
	month = jul,
	year = {2020},
	pages = {5454--5476},
}

@article{mussgnug_predictive_2022,
	title = {The predictive reframing of machine learning applications: good predictions and bad measurements},
	volume = {12},
	issn = {1879-4912, 1879-4920},
	shorttitle = {The predictive reframing of machine learning applications},
	url = {https://link.springer.com/10.1007/s13194-022-00484-8},
	doi = {10.1007/s13194-022-00484-8},
	language = {en},
	number = {3},
	urldate = {2024-04-16},
	journal = {European Journal for Philosophy of Science},
	author = {Mussgnug, Alexander Martin},
	month = sep,
	year = {2022},
	pages = {55},
}

@inproceedings{raji_fallacy_2022,
	address = {New York, NY, USA},
	series = {{FAccT} '22},
	title = {The {Fallacy} of {AI} {Functionality}},
	isbn = {9781450393522},
	url = {https://dl.acm.org/doi/10.1145/3531146.3533158},
	doi = {10.1145/3531146.3533158},
	urldate = {2023-03-20},
	booktitle = {2022 {ACM} {Conference} on {Fairness}, {Accountability}, and {Transparency}},
	publisher = {Association for Computing Machinery},
	author = {Raji, Inioluwa Deborah and Kumar, I. Elizabeth and Horowitz, Aaron and Selbst, Andrew},
	month = jun,
	year = {2022},
	pages = {959--972},
}

@inproceedings{raji_closing_2020,
	address = {New York, NY, USA},
	series = {{FAT}* '20},
	title = {Closing the {AI} accountability gap: defining an end-to-end framework for internal algorithmic auditing},
	isbn = {9781450369367},
	shorttitle = {Closing the {AI} accountability gap},
	url = {https://dl.acm.org/doi/10.1145/3351095.3372873},
	doi = {10.1145/3351095.3372873},
	urldate = {2023-03-19},
	booktitle = {Proceedings of the 2020 {Conference} on {Fairness}, {Accountability}, and {Transparency}},
	publisher = {Association for Computing Machinery},
	author = {Raji, Inioluwa Deborah and Smart, Andrew and White, Rebecca N. and Mitchell, Margaret and Gebru, Timnit and Hutchinson, Ben and Smith-Loud, Jamila and Theron, Daniel and Barnes, Parker},
	month = jan,
	year = {2020},
	pages = {33--44},
}

@inproceedings{jacobs_measurement_2021,
	address = {New York, NY, USA},
	series = {{FAccT}},
	title = {Measurement and {Fairness}},
	isbn = {978-1-4503-8309-7},
	url = {https://dl.acm.org/doi/10.1145/3442188.3445901},
	doi = {10.1145/3442188.3445901},
	urldate = {2023-04-06},
	booktitle = {Proceedings of the 2021 {ACM} {Conference} on {Fairness}, {Accountability}, and {Transparency}},
	publisher = {Association for Computing Machinery},
	author = {Jacobs, Abigail Z. and Wallach, Hanna},
	month = mar,
	year = {2021},
	pages = {375--385},
}

@inproceedings{wilson_building_2021,
author = {Wilson, Christo and Ghosh, Avijit and Jiang, Shan and Mislove, Alan and Baker, Lewis and Szary, Janelle and Trindel, Kelly and Polli, Frida},
title = {Building and Auditing Fair Algorithms: A Case Study in Candidate Screening},
year = {2021},
isbn = {9781450383097},
publisher = {Association for Computing Machinery},
address = {New York, NY, USA},
url = {https://doi.org/10.1145/3442188.3445928},
doi = {10.1145/3442188.3445928},
booktitle = {Proceedings of the 2021 ACM Conference on Fairness, Accountability, and Transparency},
pages = {666–677},
numpages = {12},
location = {Virtual Event, Canada},
series = {FAccT '21}
}

@misc{sloane_algorithmic_2021,
	title = {The {Algorithmic} {Auditing} {Trap}},
	url = {https://onezero.medium.com/the-algorithmic-auditing-trap-9a6f{2D}4d461d},
	language = {en},
	urldate = {2021-04-13},
	journal = {Medium OneZero},
	author = {Sloane, Mona},
	month = mar,
	year = {2021},
}

@inproceedings{rhea_resume_2022,
	address = {New York, NY, USA},
	series = {{AIES}},
	title = {Resume {Format}, {LinkedIn} {URLs} and {Other} {Unexpected} {Influences} on {AI} {Personality} {Prediction} in {Hiring}: {Results} of an {Audit}},
	isbn = {978-1-4503-9247-1},
	shorttitle = {Resume {Format}, {LinkedIn} {URLs} and {Other} {Unexpected} {Influences} on {AI} {Personality} {Prediction} in {Hiring}},
	url = {https://dl.acm.org/doi/10.1145/3514094.3534189},
	doi = {10.1145/3514094.3534189},
	urldate = {2023-04-06},
	booktitle = {Proceedings of the 2022 {AAAI}/{ACM} {Conference} on {AI}, {Ethics}, and {Society}},
	publisher = {Association for Computing Machinery},
	author = {Rhea, Alene and Markey, Kelsey and D'Arinzo, Lauren and Schellmann, Hilke and Sloane, Mona and Squires, Paul and Stoyanovich, Julia},
	month = jul,
	year = {2022},
	pages = {572--587},
}

@book{shove_dynamics_2012,
	address = {Los Angeles},
	title = {The dynamics of social practice: everyday life and how it changes},
	isbn = {9780857020420 9780857020437},
	shorttitle = {The dynamics of social practice},
	publisher = {SAGE},
	author = {Shove, Elizabeth and Pantzar, Mika and Watson, Matt},
	year = {2012},
	note = {OCLC: ocn783129226},
}

@book{shove_design_2007,
	address = {New York, NY},
	series = {Cultures of consumption series},
	title = {The design of everyday life},
	isbn = {9781845206826 9781845206833},
	publisher = {Berg},
	editor = {Shove, Elizabeth},
	year = {2007},
}

@article{rhea2022external,
  title={An external stability audit framework to test the validity of personality prediction in AI hiring},
  author={Rhea, Alene K and Markey, Kelsey and D’Arinzo, Lauren and Schellmann, Hilke and Sloane, Mona and Squires, Paul and Arif Khan, Falaah and Stoyanovich, Julia},
  journal={Data Mining and Knowledge Discovery},
  volume={36},
  number={6},
  pages={2153--2193},
  year={2022},
  publisher={Springer}
}

@inproceedings{bodenheimer1997process,
  title={The process of motion capture: Dealing with the data},
  author={Bodenheimer, Bobby and Rose, Chuck and Rosenthal, Seth and Pella, John},
  booktitle={Computer Animation and Simulation’97: Proceedings of the Eurographics Workshop in Budapest, Hungary, September 2--3, 1997},
  pages={3--18},
  year={1997},
  organization={Springer}
}

@article{mokander2023auditing,
  title={Auditing of AI: Legal, ethical and technical approaches},
  author={M{\"o}kander, Jakob},
  journal={Digital Society},
  volume={2},
  number={3},
  pages={49},
  year={2023},
  publisher={Springer}
}

@article{mundermann2006evolution,
  title={The evolution of methods for the capture of human movement leading to markerless motion capture for biomechanical applications},
  author={M{\"u}ndermann, Lars and Corazza, Stefano and Andriacchi, Thomas P},
  journal={Journal of neuroengineering and rehabilitation},
  volume={3},
  number={1},
  pages={6},
  year={2006},
  publisher={Springer}
}

@article{berner2020concurrent,
  title={Concurrent validity and within-session reliability of gait kinematics measured using an inertial motion capture system with repeated calibration},
  author={Berner, Karina and Cockcroft, John and Morris, Linzette D and Louw, Quinette},
  journal={Journal of Bodywork and Movement Therapies},
  volume={24},
  number={4},
  pages={251--260},
  year={2020},
  publisher={Elsevier}
}

@article{benjamini2000adaptive,
  title={On the adaptive control of the false discovery rate in multiple testing with independent statistics},
  author={Benjamini, Yoav and Hochberg, Yosef},
  journal={Journal of educational and Behavioral Statistics},
  volume={25},
  number={1},
  pages={60--83},
  year={2000},
  publisher={Sage Publications Sage CA: Los Angeles, CA}
}

@article{dykes2009towards,
  title={Towards a new disciplinary framework for contemporary creative design practice},
  author={Dykes, Thomas H and Rodgers, Paul A and Smyth, Michael},
  journal={CoDesign},
  volume={5},
  number={2},
  pages={99--116},
  year={2009},
  publisher={Taylor \& Francis}
}

@Article{paloschi_mocap,
AUTHOR = {Paloschi, Davide and Bravi, Marco and Schena, Emiliano and Miccinilli, Sandra and Morrone, Michelangelo and Sterzi, Silvia and Saccomandi, Paola and Massaroni, Carlo},
TITLE = {Validation and Assessment of a Posture Measurement System with Magneto-Inertial Measurement Units},
JOURNAL = {Sensors},
VOLUME = {21},
YEAR = {2021},
NUMBER = {19},
ARTICLE-NUMBER = {6610},
URL = {https://www.mdpi.com/1424-8220/21/19/6610},
PubMedID = {34640930},
ISSN = {1424-8220},
DOI = {10.3390/s21196610}
}

@article{chu_mocap,
title = {Validation of a video-based motion analysis technique in 3-D dynamic scapular kinematic measurements},
journal = {Journal of Biomechanics},
volume = {45},
number = {14},
pages = {2462-2466},
year = {2012},
issn = {0021-9290},
doi = {https://doi.org/10.1016/j.jbiomech.2012.06.025},
url = {https://www.sciencedirect.com/science/article/pii/S0021929012003855},
author = {Yungchien Chu and Jon Akins and Mita Lovalekar and Scott Tashman and Scott Lephart and Timothy Sell}
}

@article{bland_altman,
 ISSN = {00390526, 14679884},
 URL = {http://www.jstor.org/stable/2987937},
 author = {D. G. Altman and J. M. Bland},
 journal = {Journal of the Royal Statistical Society. Series D (The Statistician)},
 number = {3},
 pages = {307--317},
 publisher = {[Royal Statistical Society, Wiley]},
 title = {Measurement in Medicine: The Analysis of Method Comparison Studies},
 urldate = {2026-04-23},
 volume = {32},
 year = {1983}
}

@misc{HHS,
  title   = "Plan and operation of the Third National Health and Nutrition
             Examination Survey, 1988-94. Series 1: programs and collection
             procedures",
  journal = "Vital and Health Statistics",
  vol=1, 
  issue=32,
  pages   = "1--407",
  month   =  jul,
  year    =  1994,
  url = {https://www.cdc.gov/nchs/data/series/sr_01/sr01_032.pdf},
  author={{U.S. Department of Health and Human Services Public Heaith Service Centers for Disease Control and Prevention National Center for Health Statistics}}
}

@article{sloane_introducing_2022,
title={Introducing a Practice-based Compliance Framework for Addressing New Regulatory Challenges in the AI Field}, 
author={Mona Sloane and Emanuel Moss},
year=2022,
journal={TechReg Chronicle},
month={mar},
url={https://papers.ssrn.com/sol3/papers.cfm?abstract_id=4486259}}

@inproceedings{
kirk2024the,
title={The {PRISM} Alignment Dataset: What Participatory, Representative and Individualised Human Feedback Reveals About the Subjective and Multicultural Alignment of Large Language Models},
author={Hannah Rose Kirk and Alexander Whitefield and Paul R{\"o}ttger and Andrew Michael Bean and Katerina Margatina and Rafael Mosquera and Juan Manuel Ciro and Max Bartolo and Adina Williams and He He and Bertie Vidgen and Scott A. Hale},
booktitle={The Thirty-eight Conference on Neural Information Processing Systems Datasets and Benchmarks Track},
year={2024},
url={https://openreview.net/forum?id=DFr5hteojx}
}

@inproceedings{fleisig_when_2023,
	address = {Singapore},
	series = {{EMNLP}},
	title = {When the {Majority} is {Wrong}: {Modeling} {Annotator} {Disagreement} for {Subjective} {Tasks}},
	shorttitle = {When the {Majority} is {Wrong}},
	url = {https://aclanthology.org/2023.emnlp-main.415},
	doi = {10.18653/v1/2023.emnlp-main.415},
	language = {en},
	urldate = {2025-09-22},
	booktitle = {Proceedings of the 2023 {Conference} on {Empirical} {Methods} in {Natural} {Language} {Processing}},
	publisher = {Association for Computational Linguistics},
	author = {Fleisig, Eve and Abebe, Rediet and Klein, Dan},
	year = {2023},
	pages = {6715--6726},
}

@book{hand2004measurement,
  title={Measurement Theory and Practice: The World Through Quantification},
  author={Hand, D.J.},
  isbn={9780340677834},
  lccn={2005297614},
  url={https://books.google.com/books?id=7B1gRQAACAAJ},
  year={2004},
  publisher={Wiley}
}

@misc{sloane_assessing_2023,
author = {Sloane, Mona and Moss, Emanuel},
year = {2023},
month = {01},
title = {Assessing the Assessment: Comparing Algorithmic Impact Assessments and AI Audits},
doi = {10.2139/ssrn.4486259}
}

@book{bloor1991knowledge,
  title     = "Knowledge and Social Imagery",
  author    = "Bloor, David",
  publisher = "University of Chicago Press",
  edition   =  2,
  month     =  jul,
  year      =  1991,
  address   = "Chicago, IL",
  language  = "en"
}

@article{jasanoff2019controversy,
  title={Controversy studies},
  author={Jasanoff, Sheila},
  journal={The Blackwell encyclopedia of sociology},
  pages={1--5},
  year={2019},
  publisher={John Wiley \& Sons, Ltd.}
}

@book{klein1988science,
  title={The Science of Measurement: A Historical Survey},
  author={Klein, H.A.},
  isbn={9780486258393},
  lccn={lc88025858},
  series={Dover Science Books},
  url={https://books.google.com/books?id=tYZGAAAAYAAJ},
  year={1988},
  publisher={Dover Publications}
}

@article{fuller2016embrace,
  title={Embrace the inner fox: Post-truth as the STS symmetry principle universalized},
  author={Fuller, Steve},
  journal={Social Epistemology Review and Reply Collective},
  volume={25},
  year={2016}
}

@article{maguire2009birth,
  title={The birth of biometric security},
  author={Maguire, Mark},
  journal={Anthropology today},
  volume={25},
  number={2},
  pages={9--14},
  year={2009},
  publisher={Wiley Online Library}
}

@article{keller2023skin,
  title={From skin to skeleton: Towards biomechanically accurate 3d digital humans},
  author={Keller, Marilyn and Werling, Keenon and Shin, Soyong and Delp, Scott and Pujades, Sergi and Liu, C Karen and Black, Michael J},
  journal={ACM Transactions on Graphics (TOG)},
  volume={42},
  number={6},
  pages={1--12},
  year={2023},
  publisher={ACM New York, NY, USA}
}

@article{blumrosen2016real,
  title={A real-time kinect signature-based patient home monitoring system},
  author={Blumrosen, Gaddi and Miron, Yael and Intrator, Nathan and Plotnik, Meir},
  journal={Sensors},
  volume={16},
  number={11},
  pages={1965},
  year={2016},
  publisher={MDPI}
}

@article{su2014kinect,
  title={Kinect-enabled home-based rehabilitation system using Dynamic Time Warping and fuzzy logic},
  author={Su, Chuan-Jun and Chiang, Chang-Yu and Huang, Jing-Yan},
  journal={Applied Soft Computing},
  volume={22},
  pages={652--666},
  year={2014},
  publisher={Elsevier}
}

@article{landers2022game,
  title={Game-based, gamified, and gamefully designed assessments for employee selection: Definitions, distinctions, design, and validation},
  author={Landers, Richard N and Sanchez, Diana R},
  journal={International Journal of Selection and Assessment},
  volume={30},
  number={1},
  pages={1--13},
  year={2022},
  publisher={Wiley Online Library}
}

@article{turda2010race,
  title={Race, science, and eugenics in the twentieth century},
  author={Turda, Marius},
  journal={The Oxford handbook of the history of eugenics},
  pages={62--79},
  year={2010},
  publisher={Oxford University Press Oxford}
}

@article{clever2023biometry,
  title={Biometry against fascism: Geoffrey Morant, race, and anti-racism in twentieth-century physical anthropology},
  author={Clever, Iris},
  journal={Isis},
  volume={114},
  number={1},
  pages={25--49},
  year={2023},
  publisher={The University of Chicago Press Chicago, IL}
}

@article{sloane2023introducing,
  title={Introducing contextual transparency for automated decision systems},
  author={Sloane, Mona and Solano-Kamaiko, Ian Ren{\'e} and Yuan, Jun and Dasgupta, Aritra and Stoyanovich, Julia},
  journal={Nature Machine Intelligence},
  volume={5},
  number={3},
  pages={187--195},
  year={2023},
  publisher={Nature Publishing Group UK London}
}

@Article{Jeong2023,
author={Jeong, Su-Min
and Lee, Dong Hoon
and Rezende, Leandro F. M.
and Giovannucci, Edward L.},
title={Different correlation of body mass index with body fatness and obesity-related biomarker according to age, sex and race-ethnicity},
journal={Scientific Reports},
year={2023},
month={Mar},
day={01},
volume={13},
number={1},
pages={3472},
issn={2045-2322},
doi={10.1038/s41598-023-30527-w},
url={https://doi.org/10.1038/s41598-023-30527-w}
}

@ARTICLE{Pray2023-ml,
  title     = "The history and faults of the body mass index and where to look
               next: A literature review",
  author    = "Pray, Rachel and Riskin, Suzanne",
  journal   = "Cureus",
  publisher = "Springer Science and Business Media LLC",
  volume    =  15,
  number    =  11,
  pages     = "e48230",
  month     =  nov,
  year      =  2023,
  language  = "en"
}

\appendix
\section{Anthropometric Measurement Protocol}
In Phase 1 of our study, participants underwent a manual measurement protocol. Two researchers collected anthropometric data using soft and hard tape measures and a digital scale.

\begin{itemize}
    \item \textbf{Standing height:} Participants were asked to stand with their feet shoulder-width apart and their back to a wall on which a tape measure was affixed. The primary measurer placed a straightedge on top of the participant’s head, asked the participant to breathe in, and marked a line where the straightedge touched the wall.
    \item \textbf{Wingspan:} Participants were asked to stand in a T-pose with their palms perpendicular to the ground. Wingspan was measured across the back from the tips of the participant’s middle fingers.
    \item \textbf{Sitting height:} Participants were asked to sit in a chair, which had a height of 43 cm, and keep their back perpendicular to their legs. The primary measurer placed a straightedge on top of the participant’s head, asked the participant to breathe in, and marked a line where the straightedge touched the wall.
    \item \textbf{Upper leg length (R):} Participants were asked to sit in a chair and draw a line along their right hip crease to the center of the leg. The primary measurer then recorded the distance along the outside of the thigh from the identified location to the knee joint.
    \item \textbf{Biacromial breadth:} The primary measurer asked participants to raise their arms and identified the ‘hinge’ of that motion, then measured the distance between hinges across the back.
    \item \textbf{Upper arm length (R):} Participants were asked to stand up and relax their right arm. The primary measurer measured the distance from the same shoulder ‘hinge’ identified above to the bottom of the elbow, measuring along the outside of the arm.
    \item \textbf{Upper arm circumference (R):} Participants were asked to stand up and relax their right arm. The primary measurer measured the circumference of the arm halfway between the shoulder and the elbow.
    \item \textbf{Abdominal circumference:} Participants were asked to identify the top of their hip bones. The primary measurer then measured the distance from the top of the right hip bone to the top of the left hip bone and back.
    \item \textbf{Hip circumference:} Participants were asked to find the center of the hip by drawing the same line across their hip crease as they did to identify the starting point for upper leg length. The primary measurer then measured the distance from the center of the right hip to the center of the left hip and back.
    \item \textbf{Thigh circumference (R):} Participants were asked to stand with their right foot slightly ahead of their left foot and their weight on their left foot. The primary measurer measured the circumference of the arm halfway between the center of the hip (as identified above) and the knee.
    \item \textbf{Head circumference.}
    \item \textbf{Weight:} Participants were asked to step on a digital scale. They were given the option of stepping on backwards if they did not wish to see their weight.
\end{itemize}

\paragraph{Limitations.} We note two potential limitations of our anthropometric protocol. First, the research team was not previously trained in anthropometric measurement, meaning that there is a high likelihood of inter- and intra-rater variation in measurements. Second, because we offered participants the option of being measured by a researcher of their same sex, most male participants were measured by one (male) researcher and most female participants were measured by another (female) researcher.

\section{Motion Capture Protocol}
\textbf{TECH CHECK}
\begin{itemize}
    \item Before we begin, we need to make sure all tech is up and running. 
    \item Is mocap recording? 
    \begin{itemize}
        \item \textit{Wait for confirmation from technical lead}
    \end{itemize}
    \item Is depth camera recording?
    \begin{itemize}
        \item \textit{Wait for confirmation from technical lead}
    \end{itemize}
    \item Is the screen capture recording?
    \begin{itemize}
        \item \textit{Wait for confirmation from technical lead}
    \end{itemize}
\end{itemize}
\textbf{START}
\begin{itemize}
    \item Calibration (T-Pose)
    \begin{itemize}
        \item We are going to start our movement session with a T-pose. The T-pose means you are standing with your feet shoulder width apart. Your arms are raised to 90 degrees. Your palms are facing down.
        \item GO
        \item RELEASE
    \end{itemize}
    \item WALK
    \begin{itemize}
        \item The next movement is a walk. We are asking you to walk normally from CENTER POINT to POINT A in front of you, turn around, walk all the way back to POINT B, turn around and walk back to CENTER POINT.
        \item GO
        \item RELEASE
    \end{itemize}
    \item SIT
    \begin{itemize}
        \item The next movement is sitting down. For that, we are moving a chair over to you and will place it behind you. From your standing position, please sit down on the chair and stay seated for three seconds. Your feet should be flat on the floor if they reach (otherwise they should hang). After three seconds, please stand up again. We will repeat this movement once.
        \item GO
        \item RELEASE
        \item GO
        \item RELEASE
    \end{itemize}
    \item ROTATE
    \begin{itemize}
        \item Our next movement is rotating. For this movement, we ask you to slowly rotate in place. You will be doing so from a standing position, and you will be rotating to the right. 
        \item GO
        \item RELEASE
    \end{itemize}
    \item T-POSE
    \begin{itemize}
        \item We will now do another simple T-pose. The T-pose means you are standing with your feet shoulder width apart. Your arms are raised to 90 degrees. Your palms are facing down. Please hold the T-pose for three seconds. We will repeat this movement once.
        \item GO
        \item RELEASE
        \item GO
        \item RELEASE
    \end{itemize}
    \item SQUAT
    \begin{itemize}
        \item The next movement is squatting. For this movement, please squat down in a comfortable position. Please hold the squat for 3 seconds, then return to your neutral standing position upon release.
        \item GO
        \item RELEASE
    \end{itemize}
    \item ARM SWING
    \begin{itemize}
        \item The next movement is arm swinging. For this movement, please swing your arms front to back for three seconds. 
        \item GO
        \item RELEASE
    \end{itemize}
    \item ARM CIRCLES
    \begin{itemize}
        \item The next movement is arm circles. For this movement, you will put your arms into the T-pose and rotate your arms in small circles. Do this for five seconds. 
        \item GO
        \item RELEASE
    \end{itemize}
    \item JUMP
    \begin{itemize}
        \item The next movement is jumping. From your standing position, please jump up straight into the air once. We are going to repeat this movement once. 
        \item GO
        \item RELEASE
        \item GO
        \item RELEASE
    \end{itemize}
    \item KICK
    \begin{itemize}
        \item The next movement is kicking. From your standing position, use one of your legs to pretend you are kicking a ball, swinging your leg from the back to the front.
        \item GO
        \item RELEASE
    \end{itemize}
    \item We're going to do some work with props now.
    \item CARRY KETTLEBELL (FRONT)
    \begin{itemize}
        \item The next movement is carrying a kettlebell in front of you. This kettlebell weighs 10 pounds and we ask you to carry it for a short distance. For this movement, please pick up the kettlebell with both hands and hold it in front of you, as far away from your body as is comfortable for you. Then, walk to POINT A, turn around, walk all the way back to POINT B, turn around, walk back to CENTER POINT, put the kettlebell down in front of you, and return to your neutral position. 
        \item GO
        \item RELEASE
    \end{itemize}
    \item CARRY KETTLEBELL (SIDE)
    \begin{itemize}
        \item The next movement is carrying a kettlebell on your side. For this movement, please pick up the kettlebell with your dominant hand and hold it at your side. Carry it to POINT A, turn around, carry it all the way to POINT B, turn around, walk back to CENTER POINT, put the kettlebell down in front of you, and return to your neutral position. 
        \item GO
        \item RELEASE
    \end{itemize}
    \item CARRY YOGA BALL
    \begin{itemize}
        \item The next movement is carrying a giant ball that is very light. For this movement, pick up the giant ball, carry it to POINT A, turn around, carry it all the way back to POINT B, turn around, walk back to CENTER POINT, put the giant ball down in front of you, and return to your neutral position. 
        \item GO
        \item RELEASE
    \end{itemize}
    \item STEP
    \begin{itemize}
        \item  The next movement is stepping onto a platform. For this movement, please step forward onto the platform one foot at a time so that both of your feet are on the platform, and then step backward down from it one foot at a time. Then step forward onto the platform again, and finally step forwards off it.
        \item GO
        \item RELEASE
    \end{itemize}
    \item BOUNCE BALL
    \begin{itemize}
        \item  The next movement is bouncing a ball. From your standing position, please bounce the ball to the floor and catch it with both hands. Repeat this 3 times. 
        \item GO
        \item RELEASE
    \end{itemize}
    \item CROSS TOE TOUCH
    \begin{itemize}
        \item The next movement is cross toe touching. Go into a T-pose and spread your legs as far as possible. Then, touch your left toe with your right hand, rise back up, and then touch your right toe with your left hand. Repeat this movement once. 
        \item GO
        \item RELEASE
        \item GO
        \item RELEASE
    \end{itemize}
    \item RUN IN PLACE
    \begin{itemize}
        \item The next movement is running on the spot. From your neutral position, just run on the spot for five seconds. 
        \item GO
        \item RELEASE
    \end{itemize}
    \item USE PHONE
    \begin{itemize}
        \item The next movement is using your cell phone. With your dominant hand, pretend you are dialing a number on the phone. 
        \item GO
        \item RELEASE
    \end{itemize}
    \item POINT
    \begin{itemize}
        \item The next movement is pointing at something. With one hand, point to something far away. Then, with the same hand, point in a different direction. Then, point to something far away with your other hand. 
        \item GO
        \item RELEASE
    \end{itemize}
    \item CHECK WATCH
    \begin{itemize}
        \item The next movement is to check the time on your watch. For this movement, pretend you are checking the time on your watch, then relax your arm. Repeat this movement once.
        \item GO
        \item RELEASE
    \end{itemize}
    \item STAY WARM
    \begin{itemize}
        \item The next movement is staying warm. From your standing position, just rub your hands together as if you were cold and trying to stay warm.
        \item GO
        \item RELEASE
    \end{itemize}
    \item CROSS ARMS
    \begin{itemize}
        \item  The next movement is crossing your arms. For this movement, just cross your arms in front of you and hold them for five seconds.
        \item GO
        \item RELEASE
    \end{itemize}
    \item PLANK
    \begin{itemize}
        \item The second-to-last movement is a plank on the floor. For this movement, please kneel on the ground, put your palms on the ground, extend your legs behind you, and hold for five seconds. If you prefer, you can keep your knees on the ground instead of extending your legs.
        \item GO
        \item RELEASE
    \end{itemize}
    \item T-POSE
    \begin{itemize}
        \item And we are ending in a T-pose. 
        \item GO
        \item RELEASE
    \end{itemize}
\end{itemize}

\section{Data Processing and Code Availability}
Post-session, one researcher was responsible for the initial export of skeletal measurements from the mocap files. The processing pipeline and analysis scripts are hosted in the project repository (\url{https://github.com/HaukeCornell/Audit_Skeletons}).     
\begin{itemize}
    \item \textbf{Data Structure:} The repository contains the raw manual measurement logs (\textit{Manual\_Measurement\_Data}) and the aggregated mocap metrics (\textit{mocap\_skeleton \_measurements.csv}).
    \item \textbf{Analysis:} The statistical audit, including the comparison of manual versus digital measurements, was executed using the Jupyter notebooks provided in the \textit{audit\_data} and \textit{Scripts} directories.
\end{itemize}

\section{Detailed Audit Results}
We adapt the Bland-Altman approach with nested regression models to explore the relationship between different measurement types and facets of interest. We rely on the standard $p<0.05$ criteria, corrected for multiple comparisons with the Benjamini-Hochberg method, for determining significance of regression coefficients. We  report raw p-values along with statistical significance according to the Benjamini-Hochberg correction. This correction mitigates the false discovery rate from  
multiple independent tests of related hypotheses \cite{benjamini2000adaptive}, where we are testing multiple nested (not independent) models across multiple measurement outcomes (e.g., standing height, upper arm length) and measurement types (manual and inferred by mocap).

\onecolumn
\begin{small}
\begin{longtable}{p{2cm} ll lll c}

    \caption{Summary of Regression Models and Variable Coefficients (Adj. $p$-values)}
    \label{tab:comprehensive_regression_2} \\
    \toprule
    \textbf{Dependent} & \textbf{Model} & \textbf{Variable} & {\textbf{Coef.}} & {\textbf{$p$-value}} & {\textbf{Adj. $p$-value}} & \textbf{Sig. ($\alpha=0.05$)} \\
    \midrule
    \endfirsthead

    \multicolumn{7}{c}{\tablename\ \thetable\ (\textit{Continued})} \\
    \multicolumn{7}{r}{\textit{}}\\
    \toprule
    \textbf{Dependent} & \textbf{Model} & \textbf{Variable} & {\textbf{Coef.}} & {\textbf{$p$-value}} & {\textbf{Adj. $p$-value}} & \textbf{Sig. ($\alpha=0.05$)} \\
    \midrule
    \endhead

    \multicolumn{7}{r}{\textit{}}\\
    \multicolumn{7}{r}{\textit{Continued}} \\
    \bottomrule
    \endfoot

    \bottomrule
    \endlastfoot

    \multirow{12}{*}{\shortstack[l]{\textbf{standing}\\\textbf{height}}} & {avg\_h} & avg\_standing\_height & $-0.0623$ & $0.1352$ & $0.4034$ & False \\
    \cmidrule{2-7}
    & {avg\_h + sess} & all vars 
    &   & $>0.2$ & $>0.4$ & False  \\
    \cmidrule{2-7}
    & {avg\_h + sess + BMI} & all vars &   & $>0.05$ & $>0.2$ & False \\
    \cmidrule{2-7}
    & \multirow{4}{*}{avg\_h + sess + BMI + wgt} & avg\_standing\_height & $0.7171$ & $<0.001$ & $<0.001$ & \textbf{True} \\
    & & BMI & $2.5695$ & $<0.001$ & $<0.001$ & \textbf{True} \\
    & & weight & $-0.9054$ & $<0.001$ & $<0.001$ & \textbf{True} \\
    & & session\_no\_2 & $-0.2441$ & $0.6826$ & $0.8365$ & False \\
    \cmidrule{2-7}
    & \multirow{5}{*}{avg\_h + sess + BMI + wgt + sex} & avg\_standing\_height & $0.7572$ & $<0.001$ & $<0.001$ & \textbf{True} \\
    & & BMI & $2.6575$ & $<0.001$ & $<0.001$ & \textbf{True} \\
    & & weight & $-0.9159$ & $<0.001$ & $<0.001$ & \textbf{True} \\
    & & session\_no\_2 & $-0.1989$ & $0.7407$ & $0.8567$ & False \\
    & & sex\_Male & $-0.9482$ & $0.3742$ & $0.6306$ & False \\

    \midrule
    \multirow{5}{*}{{\textbf{wingspan}}} & \multirow{1}{*}{avg\_w} & avg\_wingspan & $-0.1593$ & $0.0047$ & $0.0676$ & False \\
    \cmidrule{2-7}
    & {avg\_w + sess} & all vars & & $>0.007$ & $>0.08$ & False \\
    \cmidrule{2-7}
    & {avg\_w + sess + BMI} & all vars &  & $>0.028$ & $>0.16$ & False \\
    \cmidrule{2-7}
    & {avg\_w + sess + BMI + wgt} & all vars &   & $>0.20$ & $>0.45$ & False \\
    \cmidrule{2-7}
    & {avg\_w + sess + BMI + wgt + sex} & all vars &   & $>0.19$ & $>0.45$ & False \\
    
    \midrule
    \multirow{5}{*}{\shortstack[l]{\textbf{sitting}\\ \textbf{height}}} & \multirow{1}{*}{avg\_s} & avg\_sitting\_height & $-0.0510$ & $0.7028$ & $0.8396$ & False \\
    \cmidrule{2-7}
    & {avg\_s + sess} & all vars &   & $>0.46$ & $>0.69$ & False \\
    \cmidrule{2-7}
    & {avg\_s + sess + BMI} & all vars &  & $>0.44$ & $>0.68$ & False \\
    \cmidrule{2-7}
    & {avg\_s + sess + BMI + wgt} & all vars &  & $>0.005$ & $>0.071$ & False \\
    \cmidrule{2-7}
    & {avg\_s + sess + BMI + wgt + sex} &  all vars &  & $>0.006$ & $>0.074$ & False \\

    \midrule
    \multirow{5}{*}{\shortstack[l]{\textbf{upper}\\\textbf{leg}\\\textbf{length}}} & {avg\_ul} & avg\_upper\_leg\_length & $0.2284$ & $0.2119$ & $0.4548$ & False \\
    \cmidrule{2-7}
    & {avg\_ul + sess} & all vars  &  & $>0.179$ & $>0.44$ & False \\ 
    \cmidrule{2-7}
    & {avg\_ul + sess + BMI} &  all vars &  & $>0.06$ & $>0.24$ & False  \\
    \cmidrule{2-7}
    & {avg\_ul + sess + BMI + wgt} & all vars &  & $>0.047$ & $>0.216$  & False \\
    \cmidrule{2-7}
    & {avg\_ul + sess + BMI + wgt + sex} & all vars &  & $>0.13$  & $>0.40$ & False \\

    \midrule
    \multirow{15}{*}{\shortstack[l]{\textbf{biacromial}\\\textbf{breadth}}} & \multirow{1}{*}{avg\_bb} & avg\_biacromial\_breadth & $-1.1677$ & $<0.001$ & $<0.001$ & \textbf{True} \\
    \cmidrule{2-7}
    & \multirow{2}{*}{avg\_bb + sess} & avg\_biacromial\_breadth & $-1.1957$ & $<0.001$ & $<0.001$ & \textbf{True} \\
    & & session\_no\_2 & $-0.8301$ & $0.4255$ & $0.6658$ & False \\
    \cmidrule{2-7}
    & \multirow{3}{*}{avg\_bb + sess + BMI} & avg\_biacromial\_breadth & $-1.1614$ & $<0.001$ & $<0.001$ & \textbf{True} \\
    & & BMI & $-0.2422$ & $0.1241$ & $0.3972$ & False \\
    & & session\_no\_2 & $-0.7064$ & $0.4890$ & $0.7029$ & False \\
    \cmidrule{2-7}
    & \multirow{4}{*}{avg\_bb + sess + BMI + wgt} & avg\_biacromial\_breadth & $-1.2878$ & $<0.001$ & $0.0019$ & \textbf{True} \\
    & & BMI & $-0.4330$ & $0.2076$ & $0.4542$ & False \\
    & & weight & $0.0556$ & $0.5276$ & $0.7236$ & False \\
    & & session\_no\_2 & $-1.0518$ & $0.3688$ & $0.6306$ & False \\
    \cmidrule{2-7}
    & \multirow{5}{*}{avg\_bb + sess + BMI + wgt + sex} & avg\_biacromial\_breadth & $-1.3364$ & $<0.001$ & $0.0042$ & \textbf{True} \\
    & & BMI & $-0.4248$ & $0.2236$ & $0.4667$ & False \\
    & & weight & $0.0383$ & $0.7023$ & $0.8396$ & False \\
    & & session\_no\_2 & $-1.1678$ & $0.3417$ & $0.6150$ & False \\
    & & sex\_Male & $0.7418$ & $0.7063$ & $0.8396$ & False \\

    \midrule
    \multirow{5}{*}{\shortstack[l]{\textbf{upper}\\\textbf{arm}\\\textbf{length}}} & {avg\_ua} & avg\_upper\_arm\_length & $0.1967$ & $0.4125$ & $0.6583$ & False \\
    \cmidrule{2-7}
    & {avg\_ua + sess} & all vars &  & $>0.40$ & $>0.64$ & False \\
    \cmidrule{2-7}
    & \multirow{1}{*}{avg\_ua + sess + BMI} & all vars &   & $>0.37$ & $>0.63$ & False \\
    \cmidrule{2-7}
    & \multirow{1}{*}{avg\_ua + sess + BMI + wgt} & all vars &   & $>0.23$ & $>0.47$ & False \\
    \cmidrule{2-7}
    & \multirow{1}{*}{avg\_ua + sess + BMI + wgt + sex} & all vars &  & $>0.24$ & $>0.48$ & False \\
\end{longtable}
\end{small}
\twocolumn

\end{document}